\font\eighteenb=cmb10 at 18pt
\font\tenb=cmb10
\font\eighti=cmti10 at 8pt
\font\ten=cmr10 
\font\nine=cmr10 at 9pt
\font\sevensy=cmsy10 at 7pt
\font\eightb=cmb10 at 8pt
\font\eight=cmr10 at 8pt
\font\eightmi=cmmi10 at 8pt
\font\eightsy=cmsy10 at 8pt
\def\frac#1#2{{{#1}\over{#2}}}
\vsize=9in
\hsize=6.5in
\nopagenumbers
\input psfig.sty

\ten

\centerline{\eighteenb Quantum Stochastic Heating of a Trapped Ion}
\vskip12pt
\centerline{\tenb L. Horvath, M. J. Collett, and H. J. Carmichael}
\centerline{\eighti Department of Physics, University of Auckland, Private Bag 92019, Auckland,
New Zealand}
\centerline{\eighti h.carmichael@auckland.ac.nz}
\vskip6pt
\centerline{\tenb R. Fisher}
\centerline{\eighti Department of Chemistry, Technical University of Munich, Lichtenbergstrasse 4,
85747 Garching, Germany}
\centerline{\eighti robert.fisher@ch.tum.de}
\vskip6pt
\hsize=6in
\parindent=0.5in
\hang
{\bf Abstract:} The resonant heating of a harmonically trapped ion by a standing-wave light field
is described as a quantum stochastic process combining a coherent Schr\"odinger evolution with
Bohr-Einstein quantum jumps. Quantum and semi-quantum treatments are compared.\hfill\break
\raise1.3pt\hbox{{\sevensy \char13}\hskip-5.8pt}\raise1pt\hbox{{\nine c}\hskip3pt}{\nine 2007
Optical Society of America}
\hfill\break
{\eightb OCIS codes:}\hbox{\hskip3pt{\eight (270.2500) Fluctuations, relaxations, and noise;
(270.1670) Coherent optical
effects.}}

{
\raggedright

\parindent=0.2in
\hsize=6.5in

\vskip8pt
\noindent
{\tenb 1. Introduction}
\par
\vskip6pt
\noindent
Trapped ions have featured prominently in the study of elementary quantum systems over a period of many
years. Some of the earliest experiments demonstrated the phenomenon of electron shelving or\break quantum
jumps [1-3]. Subsequently, with improved methods of laser cooling [4-6], and through resolved-sideband Raman
cooling  [7,8] in particular, it became possible to prepare an ion in the ground state of a harmonic trap.
Thus, a broad range of elementary quantum physics was opened up to experimental study [9], including the
spectroscopy [10-12] and photon statistics [13-15] of single-atom resonance fluorescence; quantized Rabi
oscillation and engineered quantum states of motion [16-18] (through a ``phonon'' analogy\break of
the Jaynes-Cummings model); and quantum decoherence, specifically, of superpositions of motional
states [19,20]. Today, trapped ion systems continue to be developed as potential building blocks of\break
quantum information processors [21-23].

Theoretical work on the quantized motion of trapped ions has mostly considered small displacements\break from
equilibrium---the so-called Lamb-Dicke regime [24]. It is common for the ion to be manipulated in a standing-wave
optical potential [25,26], in which case the potential is expanded to lowest order in the Lamb-Dicke parameter 
$$
\eta=\frac{2\pi}{\lambda}\sqrt{\frac\hbar{2m\omega_T}},\eqno{(1)\hbox{\hskip2cm}}
$$
where $\eta$ is the ratio of the ion ground-state position uncertainty $\Delta x=\sqrt{\hbar/2m\omega_T}$ and the
wavelength of the standing wave $\lambda$ (multiplied by $2\pi$); $m$ is the ion mass and $\omega_T$ the frequency.
With the restriction to the Lamb-Dicke regime, in many situations only one, two, or possibly a few quanta of
excitation are considered [16,17,27,28].

The assumption $\eta\ll1$ is a feature of the semiclassical treatment of atomic motion in laser light as well,
where it underlies the diffusion or Fokker-Planck equation models [4,29-31]. These models assume that significant
change to the particle (ion) momentum is built up from very many individual momentum ``kicks'' (each of size
$\hbar k=h/\lambda$) with the wavepacket well-localized on the scale of $\lambda$; thus, in a harmonic trap
it is assumed that
$$
\Delta p/\hbar k=(2k\Delta x)^{-1}=(2\eta)^{-1}\gg1.\eqno{(2)\hbox{\hskip2cm}}
$$
Such treatments also adiabatically eliminate the internal degrees of freedom, requiring the distance\break traveled
in an atomic lifetime $\gamma^{-1}$ to be much less than $\lambda$. This imposes the additional restriction
$$
(\omega_T/\gamma)x_0\ll\lambda,\eqno{(3)\hbox{\hskip2cm}}
$$
where $x_0$ is the amplitude of oscillation in the trap.
Since, reasonably, $x_0>\Delta x=\eta\lambda/2\pi$, from Eqs.~(2) and (3), the recoil energy is constrained by
$$
\frac{\hbar^2k^2/2m}{\hbar\gamma}=\eta^2\omega_T/\gamma\ll2\pi\eta.\eqno{(4)\hbox{\hskip2cm}}
$$

In this paper we study the quantum motion of a trapped ion in laser light under conditions where none of the usual
limitations and restrictions hold. Our aim is to exhibit the trapped ion system as an example\break of a {\it quantum
stochastic process\/} in its fully elaborated form. Specifically, we consider an ion driven on a two-state electronic
resonance by a standing-wave laser field---i.e., the heating of an ion through resonance fluorescence in a standing
wave. Motion in one dimension is considered under the assumption of strong transverse confinement. The following
features are notable in relation to the limitations and restrictions above:
\par
\vskip6pt
\parindent=0.4in
\hang
\textindent{(i)\hskip4pt} heating rather than cooling of the ion takes place; thus, beginning in the ground state of
the trap,\break the amplitude of oscillation grows to eventually extend over many periods of the standing wave,
where energies (depending on $\eta$) as high as $10^4\hbar\omega_T-10^5\mkern2mu\hbar\omega_T$ are reached;
thus, quantum motion well outside the Lamb-Dicke regime is considered;
\par
\vskip6pt
\hang
\textindent{(ii)\hskip4pt} we focus upon the case
$$
\omega_T/\gamma=1,\eqno{(5)\hbox{\hskip2cm}}
$$
so inequality (3) is violated at relatively small amplitudes of oscillation; a crucial interplay develops\break between
the coherent evolution of the center-of-mass and internal degrees of freedom; even when a diffusion paradigm fits
(for sufficiently small $\eta$), the motion is not understandable as diffusion in a\break prescribed harmonic
plus optical potential;
\par
\vskip6pt
\hang
\textindent{(iii)\hskip4pt} strong laser excitation, with Rabi frequency
$$
\Omega/\gamma=2,\eqno{(6)\hbox{\hskip2cm}}
$$
is considered, but due to the interplay just noted the dressed-atom approach to atomic motion in laser light [32]
cannot be used; modulation of the Rabi frequency by the harmonic center-of-mass motion plays a central role;
\par
\vskip6pt
\hang
\textindent{(iv)\hskip4pt}
Lamb-Dicke parameters ranging from $\eta=0.2$ to $\eta=3.0$ are considered, thus demonstrating the\break
transition from a quasi-classical dynamic, for $\eta\ll1$, to a manifestly {\it quantum\/} stochastic dynamic
when $\eta>1$; this transition is similar to that from weak- to strong-coupling in the treatment of quantum
noise in cavity QED [33].
\par
\vskip6pt
\parindent=0.2in
The reported investigation makes use of full {\it quantum trajectory\/} simulations [34], quantizing both\break the
internal and the center-of-mass degrees of freedom. As an aid to understanding, approximate,\break {\it semi-quantum
trajectory\/} simulations are explored as well, where the center-of-mass motion is treated\break classically apart
from the inclusion of momentum ``kicks'' coordinated with the fluorescence. There have been previous quantum trajectory
(Monte Carlo wave-function [35]) simulations of atomic motion in laser light [36-38], but all, to our knowledge, were
carried out for conditions satisfying the restrictions of\break Eqs.~(2)--(4).

We begin in Sec.~2 with an overview, presenting a summary of results in the form of heating curves (mean energy
of the ion as a function of time) calculated for a series of Lamb-Dicke parameters ranging\break between $\eta=0.2$ and
$\eta=3.0$. Results calculated from semi-quantum and quantum trajectories are\break compared and a number of
differences identified for explanation in subsequent sections. The equations\break underlying the Monte-Carlo
simulations are presented in Sec.~3, where we demonstrate a surprising\break feature of the individual realizations of
the stochastic process; a stepwise evolution of the amplitude of oscillation of the ion is observed,
pointing to the existence of a series of metastable amplitudes on which the ion heating almost stops.
In Sec.~4 the origin of this evolution is identified as frequency modulation of the internal state Rabi oscillation.
The metastable amplitudes of oscillation are identified with zeros of the $J_0$ Bessel function and characterized
by the mean waiting time for fluorescent scattering plotted as a function of the amplitude of oscillation
in the trap. A phase-space diffusion model is derived that recovers\break the heating curves of Sec.~2 in the
limit of small Lamb-Dicke parameters, and quantitative differences\break between the semi-quantum and quantum diffusion
are explained. Finally, in Sec.~5, Lamb-Dicke\break parameters of order unity are considered, where differences
between a diffusion and a jump process\break are revealed. Here semi-quantum and quantum trajectory models are
qualitatively different in their\break predictions. An observed quantum suppression of the ion heating
rate is explained by the delocalization of the center-of-mass wavepacket across the standing wave.

\vskip6pt
\noindent
{\tenb 2. Quantum and semi-quantum heating rates}
\par
\vskip6pt
\noindent
We consider an ion of mass $m$ trapped harmonically in one dimension with trap frequency $\omega_T$. A classical
laser field resonantly excites a closed two-state electronic transition of the ion, with decay rate $\gamma$ and
Rabi frequency $\Omega$; parameters chosen as in Eqs.~(5) and (6). The laser field forms a standing wave
along the axis of the ion motion (wavelength $\lambda$) and the equilibrium of the potential is located at an
anti-node of the standing wave.

Figure 1 shows the growth of the mean-squared amplitude of oscillation of the ion---measured in wavelengths [see
Eqs.~(20) and (21)]---as a function of time for various Lamb-Dicke parameters $\eta$. Two separate models were used
to compute these results as ensemble averages over Monte-Carlo simulations of the ion heating through fluorescence.
For the results of Fig.~1(a), the ion center-of-mass position and momentum\break are treated as classical variables,
but with a momentum ``kick'' added when a laser photon is scattered; the scattering is simulated as a quantum
trajectory for a two-state system with a definite, though oscillating,\break position within the standing
wave [34-37,39]. For the results of Fig.~1(b), full quantum trajectory\break simulations were carried out with both the
center-of-mass and internal degrees of freedom quantized. The following points are of note and will be elaborated
upon in the following sections:
\par
\vskip6pt
\parindent=0.4in
\hang
\textindent{(i)\hskip4pt} for small Lamb-Dicke parameters both sets of results approach a limit where the heating
curve\break becomes independent of $\eta$, apart from a factor $\eta^4$ that may be absorbed in the scaling of time;
in this limit the heating is well-described by a diffusion in phase-space;
\par
\vskip6pt
\hang
\textindent{(ii)\hskip4pt} semi-quantum and quantum trajectory models disagree quantitatively in the diffusion
limit\break [compare the thinest curves in Figs.~1(a) and (b)];
\par
\vskip6pt
\hang
\textindent{(iii)\hskip4pt} larger Lamb-Dicke parameters yield an $\eta$-dependence over and above the scaling of time
by $\eta^4$; thus, differences between a quantum jump and diffusion process become explicitly apparent;
\par
\vskip6pt
\hang
\textindent{(iv)\hskip4pt} for large Lamb-Dicke parameters the two sets of curves are qualitatively different; the
semi-quantum trajectories [frame (a)] show a heating rate that increases monotonically with increasing $\eta$,
while the full quantum trajectory results [frame (b)] eventually show a dramatic reduction of the heating
rate.
\par
\vskip6pt
\parindent=0.2in
Before taking up these four themes, we first present the mathematical details of our two models in Sec.~3.
Then we explore individual realizations of the quantum stochastic heating of the ion. These\break exhibit a very 
interesting evolution of the amplitude of oscillation of the ion in the trap, an underlying structure to the
stochastic dynamics quite unanticipated on the basis of the ensemble averages displayed in Fig.~1.

}

\vbox{\vskip1cm
\centerline{\hskip1cm \psfig{figure=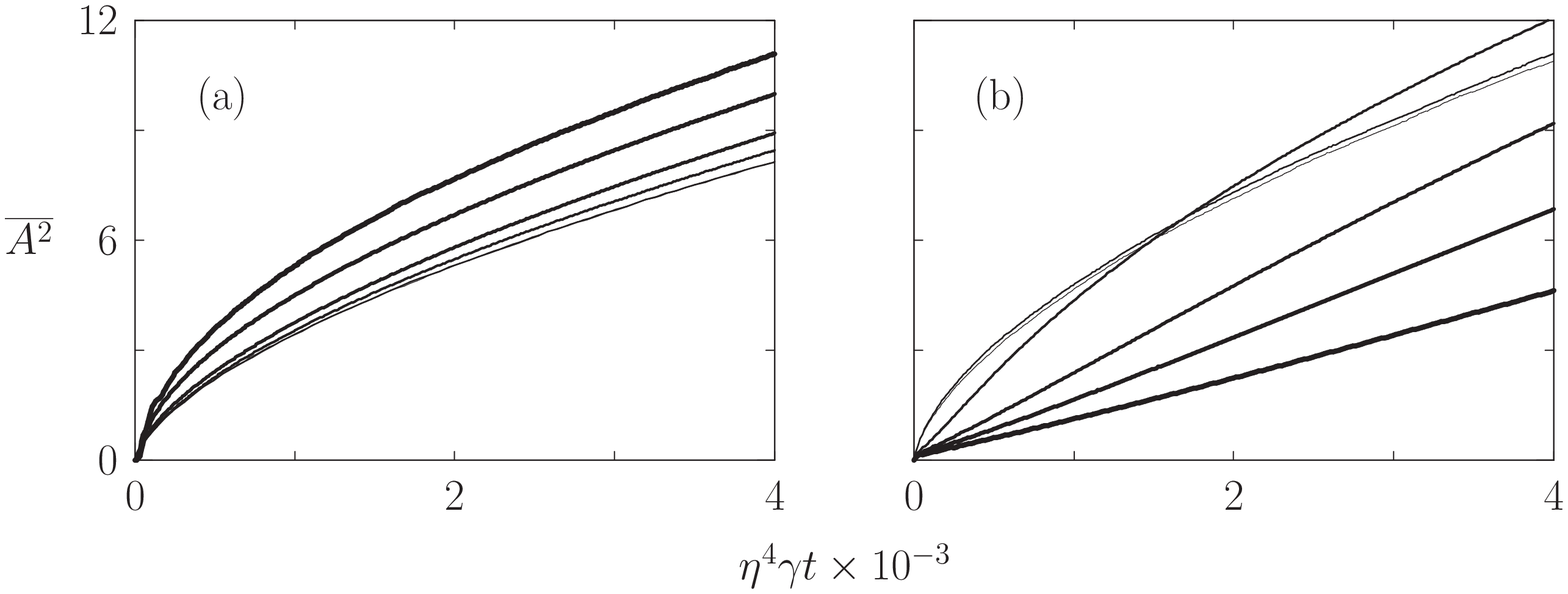,height=4.5cm}}

\vskip12pt

\hsize=5.5in
\parindent=1.0in
\baselineskip=6pt
\hang
{\eight Fig.~1. Heating of a harmonically trapped ion through resonance fluorescence in a standing-wave
field. Plots of the mean squared amplitude of oscillation as a function of time from: (a) semi-quantum
trajectory simulations---Lamb-Dicke parameters \hbox{\eightmi\char17}$\mkern3mu$
=$\mkern3mu$0.2, 0.4, 0.8, 1.4, 2.2, and 2.6 (thinest line to thickest line); (b) quantum trajectory
simulations---Lamb-Dicke parameters \hbox{\eightmi\char17}$\mkern3mu$=$\mkern3mu$0.8, 1.4, 2.2, 2.6,
2.8, and 3.0 (thinest line to thickest line).}
}

\parindent=0.2in
\hsize=6.5in

\vskip6pt

{
\raggedright

\vfill
\break
\vskip6pt
\noindent
{\tenb 3. Monte-Carlo simulations}
\par
\vskip6pt
\noindent
{\it 3.1 Semi-quantum trajectories}:
\par
\vskip6pt
\noindent
Semi-quantum trajectories treat the center-of-mass motion of the ion classically. Only the electronic states
of the ion are quantized. We represent the center-of-mass state by the point in phase-space
$$
\mkern120mu\alpha(t)=\tilde\alpha(t)e^{-i\omega_Tt}=\sqrt{\frac{m\omega_T}{2\hbar}}\left[x(t)+i\frac{p(t)}
{m\omega_T}\right],\eqno{(7)\hbox{\hskip2cm}}
$$
where $x(t)$ denotes the oscillator position and $p(t)$ its momentum. The trajectory equations are\break formulated
in a rotating frame, with $\tilde\alpha(t)$ denoting the complex amplitude of the oscillation. Following\break 
the standard quantum trajectory treatment of resonance fluorescence (see Sec.~8.2 of Ref.~[34]), the\break
conditional internal state of the ion is expanded in the interaction picture as
$$
|\bar\psi_{\rm REC}(t)\rangle=\bar c^{(-)}(t)|-\rangle+\bar c^{(+)}(t)|+\rangle,\eqno{(8)\hbox{\hskip2cm}}
$$
where $|-\rangle$ and $|+\rangle$ are, respectively, the ground and excited electronic states, and the label
${\rm REC}$ refers\break to the conditioning of the state on the scattering record prior to time $t$; ket
vectors and Schr\"odinger\break amplitudes with an overbar, as in Eq.~(8), are unnormalized. Under the assumed
resonant excitation,\break in between photon scattering events, the (unnormalized) Schr\"odinger amplitudes
obey the equations of motion
$$
\eqalignno{
\frac{d\bar c^{(-)}}{dt}&=-(\Omega/2)\cos\mkern-2mu\left[\eta\mkern-2mu\left(\tilde\alpha e^{-i\omega_Tt}
+\tilde\alpha^*e^{i\omega_Tt}\right)\right]\bar c^{(+)},\cr
\noalign{\vskip2pt}
\frac{d\bar c^{(+)}}{dt}&=(\Omega/2)\cos\mkern-2mu\left[\eta\mkern-2mu\left(\tilde\alpha e^{-i\omega_Tt}
+\tilde\alpha^*e^{i\omega_Tt}\right)\right]\bar c^{(-)}-\frac\gamma2\bar c^{(+)},&{(9)\hbox{\hskip2cm}}
}
$$
where, importantly, the Rabi frequency $\Omega$ is modulated by the harmonic motion of the ion through the\break standing
wave---using Eqs.~(1) and (7), $\cos[kx(t)]=\cos\{\eta[\alpha(t)+\alpha^*(t)]\}$. Photon scattering is accounted for
through quantum jumps, which interrupt this coherent evolution. The jumps occur at random times, with instantaneous rate
$$
R_{\rm jump}(t)=\gamma\frac{|\bar c^{(+)}(t)|^2}{|\bar c^{(-)}(t)|^2+|\bar c^{(+)}(t)|^2}.\eqno{(10)\hbox{\hskip2cm}}
$$
Each jump returns the ion to the ground electronic state,
$$
\bar c_-\to\bar c_+,\qquad\bar c_+\to0,\eqno{(11{\rm a})\hbox{\hskip2cm}\mkern-8mu}
$$
and at the same time we add a momentum ``kick'' to the center-of-mass. The magnitude of this ``kick'' is determined
by projecting the recoil for photon scattering in a randomly chosen direction $(\theta,\phi)$ onto the line of the
ion motion; thus, for photon scattering at time $t$, we implement a momentum ``kick'' in the\break rotating frame
through the semi-quantum jump
$$
\tilde\alpha\to\tilde\alpha+i\eta\sin\theta\cos\phi e^{i\omega_Tt},\eqno{(11{\rm b})\hbox{\hskip2cm}\mkern-8mu}
$$
which follows from the magnitude of the recoil $\hbar k$ and Eqs.~(1) and (7). The angle $\phi$ is uniformly\break
distributed between $0$ and $2\pi$, while $\theta$ is distributed according to the dipole radiation pattern
(oriented perpendicular to the ion motion).

A Monte-Carlo simulation must decide at each time step whether to propagate the internal state\break according to
Eq.~(9), leaving $\tilde\alpha$ unchanged, or to implement the quantum and semi-quantum jumps,\break Eqs.~(11a)
and (11b); the branching ratio is given by the jump probability $R_{\rm jump}\Delta t$. When a jump\break (photon
scattering) does occur, two random numbers determine the angles $\theta$ and $\phi$. The heating of the\break
ion results from the accumulated momentum ``kicks'' (11b). The mechanism is therefore, in principle, straightforward.
It is important to note, however, that the ``kick'' rate is not a fixed function of either space or time, but itself
evolves stochastically according to Eqs.~(9) and (10). This gives rise to an\break unanticipated complexity in the ion motion,
as illustrated in Sec.~3.4.
\par
\vskip6pt
\noindent
{\it 3.2 Quantum trajectories}:
\par
\vskip6pt
\noindent
The full quantum trajectory treatment of the ion heating is formulated as a natural extension of Eqs.~(8)--(11b).
The principle difference is that Schr\"odinger amplitudes $\bar c^{(-)}(t)$ and $\bar c^{(+)}(t)$ are replaced
by center-of-mass ket vectors, with the expansion of the total quantum state, quantized internal {\it and\/}
center-of-mass degrees of freedom,
$$
\mkern90mu|\bar\psi_{\rm REC}(t)\rangle=|\bar\psi_{\rm REC}^{(-)}(t)\rangle|-\rangle+
|\bar\psi_{\rm REC}^{(+)}(t)\rangle|+\rangle.\eqno{(12)\hbox{\hskip2cm}}
$$
The coupled equations of motion (9) for Schr\"odinger amplitudes are replaced by coupled equations of\break
motion for the (unnormalized) center-of-mass kets,
$$
\eqalignno{
\frac{d|\bar\psi_{\rm REC}^{(-)}\rangle}{dt}&=-(\Omega/2)\cos\mkern-2mu\left[\eta\mkern-2mu
\left(\hat a e^{-i\omega_Tt}+\hat a^\dagger e^{i\omega_Tt}\right)\right]|\bar\psi_{\rm REC}^{(+)}\rangle,\cr
\noalign{\vskip2pt}
\frac{d|\bar\psi_{\rm REC}^{(+)}\rangle}{dt}&=(\Omega/2)\cos\mkern-2mu\left[\eta\mkern-2mu
\left(\hat a e^{-i\omega_Tt}+\hat a^\dagger e^{i\omega_Tt}\right)\right]|\bar\psi_{\rm REC}^{(-)}\rangle
-\frac\gamma2|\bar\psi_{\rm REC}^{(+)}\rangle,&{(13)\hbox{\hskip2cm}}
}
$$
where $\hat a$ and $\hat a^\dagger$ are oscillator annihilation and creation operators, and replace the
classical phase-space variables $\tilde\alpha$ and $\tilde\alpha^*$. In line with the rotating frame
of reference used in Eqs.~(7) and (9), the interaction picture of the harmonic oscillator has been adopted
in Eqs.~(12) and (13). The jump (photon scattering) rate is calculated from the center-of-mass ket norms
in a straightforward generalization of Eq.~(10),
$$
\mkern90mu R_{\rm jump}(t)=\gamma\frac{\langle\bar\psi_{\rm REC}^{(+)}(t)|\bar\psi_{\rm REC}^{(+)}(t)\rangle}
{\langle\bar\psi_{\rm REC}^{(-)}(t)|\bar\psi_{\rm REC}^{(-)}(t)\rangle+\langle\bar\psi_{\rm REC}^{(+)}
(t)|\bar\psi_{\rm REC}^{(+)}(t)\rangle},\eqno{(14)\hbox{\hskip2cm}}
$$
and (11a) and (11b) are combined in the quantum jump
$$
\mkern90mu|\bar\psi_{\rm REC}\rangle\to\hat D\mkern-2mu\left(i\eta\sin\theta\cos\phi e^{i\omega_Tt}\right)
|\bar\psi_{\rm REC}^{(+)}\rangle|-\rangle,\eqno{(15)\hbox{\hskip2cm}}
$$
where $\hat D(\xi)=\exp(\xi\hat a^\dagger-\xi^*\hat a)$ is the displacement operator. 

Monte-Carlo simulations may be carried out on the basis of Eqs.~(13)--(15) in essentially the same\break
manner as before. The numerical demands are significantly higher, though, due to the much larger\break
number basis states involved. The two center-of-mass kets are expanded in the energy representation
of the harmonic oscillator, and we have, in fact, carried out simulations where the energy of the
center-of-mass motion rises to be of the order of $10^4\hbar\omega_T-10^5\hbar\omega_T$. This is achieved
by carrying out the\break calculations in a local frame that tracks the average complex amplitude
of the oscillation, $\langle\hat a(t)\rangle_{\rm REC}$,\break reached at the conclusion of a quantum jump.
This strategy also circumvents the need to explicitly\break implement the displacement in Eq.~(15).
\par
\vskip6pt
\noindent
{\it 3.3 Quantum trajectories in the local frame}:
\par
\vskip6pt
\noindent
We absorb the mean displacement of the center-of-mass kets from one quantum jump to another into a
change of the frame of reference. Thus, we work with the displaced center-of-mass kets,
$$
|\bar\psi_{\rm local}^{(\pm)}(t)\rangle=\hat D[-\tilde\alpha(t)]|\bar\psi_{\rm REC}^{(\pm)}(t)\rangle,
\eqno{(16)\hbox{\hskip2cm}}
$$
where $\hat D[-\tilde\alpha(t)]$ displaces the center-of-mass state to a ``local frame''. The local
frame changes in time,\break discontinuously, at the times of the quantum jumps. When it has been decided
that a photon scattering is to take place and the associated quantum jump (15) is to be executed, a displacement
of the post-jump kets is made to first cancel the displacement of the momentum ``kick'' [Eq.~(15)],
and second to move the mean complex amplitude of the oscillator back to zero---i.e., to off-set any change in
the mean brought about by the Schr\"odinger evolution during the interval that has elapsed since the
last quantum jump. Then in place of Eqs.~(13) we have equations of motion for the ket vectors in the
local frame,
$$
\eqalignno{
\frac{d|\bar\psi_{\rm local}^{(-)}\rangle}{dt}&=-(\Omega/2)\cos\mkern-2mu\left\{\eta\mkern-2mu
\left[(\tilde\alpha+\hat a)e^{-i\omega_Tt}+(\tilde\alpha^*+\hat a^\dagger)e^{i\omega_Tt}\right]\right\}
|\bar\psi_{\rm local}^{(+)}\rangle,\cr
\noalign{\vskip2pt}
\frac{d|\bar\psi_{\rm local}^{(+)}\rangle}{dt}&=(\Omega/2)\cos\mkern-2mu\left\{\eta\mkern-2mu
\left[(\tilde\alpha+\hat a)e^{-i\omega_Tt}+(\tilde\alpha^*+\hat a^\dagger)e^{i\omega_Tt}\right]\right\}
|\bar\psi_{\rm local}^{(-)}\rangle-\frac\gamma2|\bar\psi_{\rm local}^{(+)}\rangle,\cr
\noalign{\vskip2pt}
{}&{}&{(17)\hbox{\hskip2cm}}
}
$$
with jump rate
$$
\mkern100mu R_{\rm jump}(t)=\gamma\frac{\langle\bar\psi_{\rm local}^{(+)}(t)|\bar\psi_{\rm local}^{(+)}(t)\rangle}
{\langle\bar\psi_{\rm local}^{(-)}(t)|\bar\psi_{\rm local}^{(-)}(t)\rangle+\langle\bar\psi_{\rm localC}^{(+)}
(t)|\bar\psi_{\rm local}^{(+)}(t)\rangle}.\eqno{(18)\hbox{\hskip2cm}}
$$
The quantum jump is now executed by setting the mean oscillator amplitude with respect to the local frame
to zero (and setting the ion in the ground electronic state), with
$$
\mkern100mu|\bar\psi_{\rm local}^{(-)}\rangle\to\hat D(-\langle\hat a\rangle_{\rm local})|\bar\psi_{\rm local}^{(+)}\rangle,
\qquad|\bar\psi_{\rm local}^{(+)}\rangle\to0,\eqno{(19{\rm a})\hbox{\hskip2cm}\mkern-8mu}
$$
and moving the location of the local relative to the global frame, with
$$
\tilde\alpha\to\tilde\alpha+\langle\hat a\rangle_{\rm local}+i\eta\sin\theta\cos\phi e^{i\omega_Tt}.
\eqno{(19{\rm b})\hbox{\hskip2cm}\mkern-8mu}
$$
In these expression $\langle\hat a\rangle_{\rm local}$ is the conditional oscillator amplitude
expectation calculated in the pre-jump local frame.
\par
\vskip6pt
\noindent
{\it 3.4 Sample results}:
\par
\vskip6pt
\noindent
Figure~2 presents examples of the simulated trajectories for four different values of the Lamb-Dicke\break parameter,
increasing from $\eta=0.2$ in frame (a) to $\eta=2.2$ in frame (d). We plot the amplitude of\break oscillation, measured
in optical wavelengths, against time in atomic lifetimes scaled by $\eta^4$ (see Sec.~4.2). Explicitly, for semi-quantum
trajectories, using Eqs.~(1) and (7), the amplitude is given by
$$
A=\frac1\lambda\sqrt{\frac{2\hbar}{m\omega_T}}|\tilde\alpha|=\frac\eta\pi|\tilde\alpha|,
\eqno{(20)\hbox{\hskip2cm}}
$$
while for quantum trajectories
$$
\mkern100mu A=\frac\eta\pi\sqrt{\langle\hat a^\dagger\hat a\rangle_{\rm REC}}
=\frac\eta\pi\sqrt{\langle(\tilde\alpha^*+\hat a^\dagger)(\tilde\alpha+\hat a)\rangle_{\rm local}}.
\eqno{(21)\hbox{\hskip2cm}}
$$
There are two principal points of interest to be noted in the figure. First, frames (a) and (b) shown a clear
step-like evolution, which is at least qualitatively similar in the semi-quantum and quantum simulations. Long
periods of almost no heating at all (metastable amplitudes) are present, interspersed with periods of much more
rapid heating, thus making up the steps. Quantitatively, periods of essentially no heating are observed to last
as long as $10^6$ lifetimes, while the metastable amplitudes are spaced approximately $\lambda/2$ apart---at
$A=0.38,0.88,1.38,$ etc.. Second, the similarity between semi-quantum and quantum\break trajectories seems to be
breaking down in fames (c) and (d). The quantum trajectory shows no clear steps in frame (d), while the stepping
of the semi-quantum trajectory no longer keeps to the metastable\break amplitudes identified before. The task
of the following sections is to explain these features along with points (i)--(iv) of Sec.~2.

}

\vbox{\vskip0.5cm
\centerline{\psfig{figure=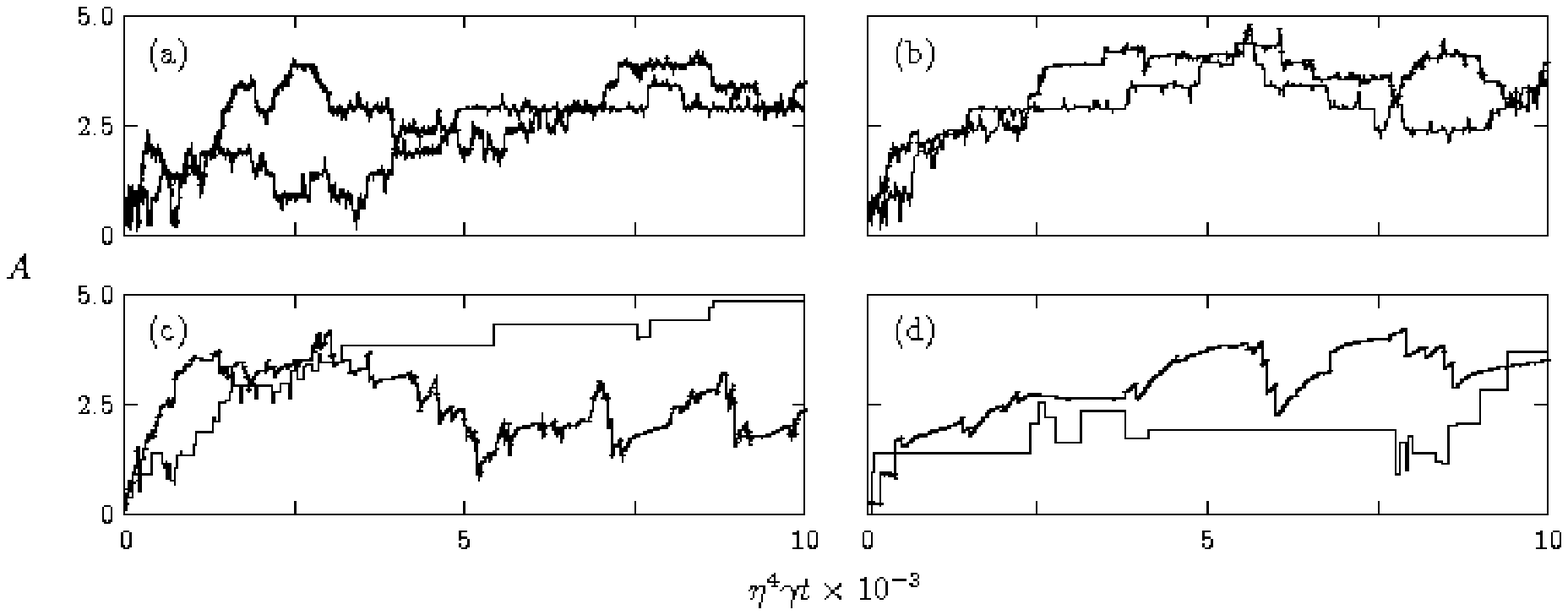,height=5.5cm}}

\vskip6pt

\hsize=6in
\parindent=0.5in
\baselineskip=6pt
\hang
{\eight Fig.~2. Sample quantum (thick curves) and semi-quantum (thin curves) trajectory simulations of the
amplitude of oscillation in the trap; for Lamb-Dicke parameters \hbox{\eightmi\char17}$\mkern3mu$
=$\mkern3mu$0.2 (a), 0.8 (b), 1.4 (c), and 2.2 (d).}
}

\parindent=0.2in
\hsize=6.5in

\vskip12pt

{
\raggedright

\vskip6pt
\noindent
{\tenb 4. Mean waiting times and the diffusion limit}
\par
\vskip6pt
\noindent
\noindent{\it 4.1 Semi-quantum waiting times}:
\par
\vskip6pt
\noindent
An explanation of the step-like evolution of Figs.~2(a) and (b) is provided by Fig.~3. Here, in the upper series of
pictures, we plot the induced Rabi oscillation and simulated photon scattering sequence for four different amplitudes
of oscillation in the trap. For simplicity, only Semi-quantum trajectories are considered in this section. Frames (a)
and (c) show the ion being fully excited, while (b) and (d) show much lower levels of excitation. The corresponding
numbers of scattered photons are high---frames (a) and (c)---and much lower---frames (b) and (d). Evidently, the
response of the ion to the modulated driving field [Eqs.~(9)],
$$
\mkern100mu(\Omega/2)\cos[kx(t)]=(\Omega/2)\cos\{2\pi A\cos[\omega_T t-\arg(\tilde\alpha)]\},\eqno{(22)\hbox{\hskip2cm}}
$$
depends strongly on the amplitude of oscillation in the trap. The observation is readily understood from the
time-averaged Rabi frequency, $\Omega J_0(2\pi A)$, which is zero at the zeros of the $J_0$ Bessel function.
The first two of these zeros occur at the amplitudes $A=0.38$ and $A=0.88$ of frames (b) and (d) in the figure,\break
while amplitudes $A=0.0$ and $A=0.6$ [frames (a) and (c)] correspond to local maxima of the Bessel\break
function. Thus, the metastable amplitudes are located by the zeros of $J_0(2\pi A)$, where the ion is only\break weakly excited
and the fluorescence rate drops. Of course, some photon scattering occurs at these zeros, since the actual response
is to the modulated driving and not to its time average; the amount decreases as $A$ increases, as seen in the comparison
between frames (b) and (d).

}

\vbox{\vskip0.5cm
\centerline{\psfig{figure=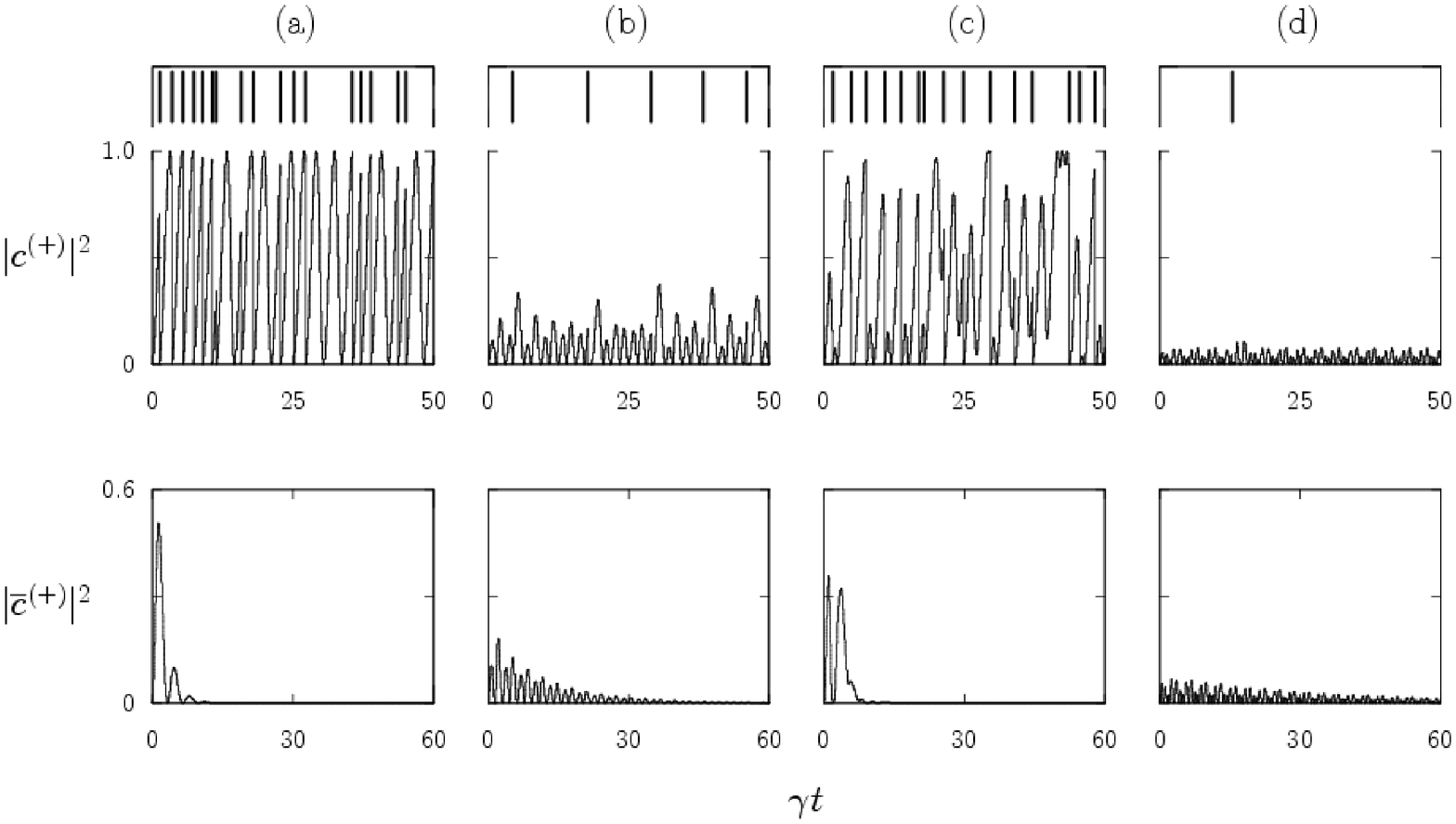,height=7.5cm}}

\vskip6pt

\hsize=6in
\parindent=0.5in
\baselineskip=6pt
\hang
{\eight Fig.~3. Semi-quantum Rabi oscillation and photon scattering sequence (upper frames) and corresponding
waiting-time distribution (lower frames) for four different amplitudes of oscillation in the trap: \hbox{\eightmi\char65}
$\mkern3mu$=$\mkern3mu$0.0 (a), 0.38 (b), 0.6 (c), and 0.88 (d).}
}

\parindent=0.2in
\hsize=6.5in

\vskip12pt

{
\raggedright

A useful way to quantify the behavior as a function of $A$ is through the distribution of times waited\break between two
successive scattering events (quantum jumps)---the so-called waiting-time distribution. This distribution is plotted for
semi-quantum trajectories in the lower series of pictures in Fig.~3, where it is narrow (width of order $\gamma\tau=1$) in
frames (a) and (c), significantly broadened in frame (b), and broader still in frame (d). The distribution depends, most
generally, on both the amplitude, $A$, and phase,
$$
\zeta=\omega_Tt-\arg(\tilde\alpha),\eqno{(23)\hbox{\hskip2cm}},
$$
of the oscillation in the trap at the time of the quantum jump that begins the interval waited $\tau$. The\break quantity
plotted in Fig.~3 has been averaged over $\zeta$. For given amplitude and phase, the waiting-time\break distribution is
computed within quantum trajectory theory as [39]
$$
W(\tau;A,\zeta)=\gamma|\bar c^{(+)}(t+\tau)|^2,\eqno{(24)\hbox{\hskip2cm}}
$$
where $\bar c^{(+)}(t+\tau)$ satisfies Eqs.~(9) with initial condition $\bar c^{(+)}(t)=0$, $\bar c^{(-)}(t)=1$.
Its mean,
$$
\bar\tau(A,\zeta)=\int_0^\infty\mkern-3mu d\tau\mkern2mu\tau W(\tau;A,\zeta),\eqno{(25)\hbox{\hskip2cm}}
$$
is sufficient to characterize the metastable states. For semi-quantum trajectories, an approximate\break analytical
expression for this mean can be derived as follows.

The task is to solve Eqs.~(9) with the initial internal state the electronic ground state. We first make a transformation
to consider the complex variable
$$
\mkern100mu\bar c^{(-)}(t+\tau)+i\bar c^{(+)}(t+\tau)=B(\tau;A,\zeta)e^{i\beta(\tau;A,\zeta)},\eqno{(26)\hbox{\hskip2cm}}
$$
where $B(\tau;A,\zeta)$ and $\beta(\tau;A,\zeta)$ are real functions of $\tau$. The equations of motion for Schr\"odinger\break
amplitudes, Eqs.~(9), then yield a solution in the form
$$
\mkern100mu B(\tau;A,\zeta)=\exp\mkern-3mu\left[-\frac\gamma2\mkern-2mu\int_0^\tau\mkern-2mu d\tau^\prime
\sin^2\mkern-2mu\beta(\tau^\prime;A,\zeta)\right],\eqno{(27)\hbox{\hskip2cm}}
$$
with the phase $\beta(\tau;A,\zeta)$ satisfying the equation
$$
\mkern100mu\frac{d\beta(\tau;A,\zeta)}{d\tau}=-(\Omega/2)\cos\left[2\pi A\cos(\omega_T\tau+\zeta)\right]
-\frac\gamma4\sin\mkern-2mu\beta(\tau;A,\zeta).\eqno{(28)\hbox{\hskip2cm}}
$$
From Eqs.~(24), (26), and (27), the waiting-time distribution is
$$
W(\tau;A,\zeta)=-\frac{d}{d\tau}B^2(\tau;A,\zeta),\eqno{(29)\hbox{\hskip2cm}}
$$
with mean waiting time, integrating Eq.~(25) by parts,
$$
\bar\tau(A,\zeta)=\int_0^\infty\mkern-2mu d\tau B^2(\tau;A,\zeta).\eqno{(30)\hbox{\hskip2cm}}
$$
The development up to this point is exact. We now introduce an approximation which holds good\break whenever the
mean waiting time is much larger than $\gamma^{-1}$. Assuming $\beta(\tau;A,\zeta)$ is very small, such that
$\sin\beta(\tau;A,\zeta)\approx\beta(\tau;A,\zeta)$, Eq.~(27) becomes
$$
\mkern100mu B(\tau;A,\zeta)=\exp\mkern-3mu\left[-\frac\gamma2\mkern-2mu\int_0^\tau\mkern-2mu d\tau^\prime
\beta^2(\tau^\prime;A,\zeta)\right],\eqno{(31)\hbox{\hskip2cm}}
$$
while Eq.~(28) can be solved for
$$
\mkern100mu\beta(\tau;A,\zeta)=(\Omega/2){\rm Re}\mkern-3mu\left[\sum_{n=-\infty}^\infty(-i)^n\frac{J_n(2\pi A)}
{\gamma/2+in\omega_T}\mkern3mue^{in(\omega_T\tau+\zeta)}\right],\eqno{(32)\hbox{\hskip2cm}}
$$
where we have used the standard Fourier series expansion of $\cos\left[2\pi A\cos(\omega_T\tau+\zeta)\right]$ [40].
We then replace $\beta^2(\tau;A,\zeta)$ in Eq.~(31) by its d.c. component (which is independent of the phase $\zeta$),
and from Eqs.~(29)--(31), the waiting-time distribution is approximated by the exponential distribution
$$
W(\tau;A)=\bar\tau^{-1}e^{-\tau/\bar\tau(A)},\eqno{(33)\hbox{\hskip2cm}}
$$
with mean waiting time
$$
\mkern100mu\gamma\bar\tau(A)=\left[(\Omega/2)^2\sum_{n=0}^\infty\frac{J_{2n}^2(2\pi A)+J_{-2n}^2(2\pi A)}
{(\gamma/2)^2+(2n\omega_T)^2}\right]^{-1}.\eqno{(34)\hbox{\hskip2cm}}
$$

Note how the dominant term in the series expansion of $\gamma\bar\tau$ vanishes for the identified metastable\break
amplitudes---when $J_0(2\pi A)=0$. The approximation (34) is compared with the exact result computed from a semi-quantum
trajectory average in Fig.~4. The two results are in remarkably good agreement\break everywhere except for amplitudes $A<1$
and in a narrow range of amplitudes in between the peaks.\break We see that with increasing center-of-mass energy, the rate of
photon scattering on the metastable\break amplitudes very quickly drops by more than two orders of magnitude. Clearly this
interplay of the\break coherent oscillation of the ion in the trap and the induced Rabi oscillation must be the principal\break
determinant of the heating curves---for small Lamb-Dicke parameters at least---and of the noted\break differences between
semi-quantum and quantum models (Fig.~1).

}

\vbox{\vskip0.75cm
\centerline{\hskip-0.5cm\psfig{figure=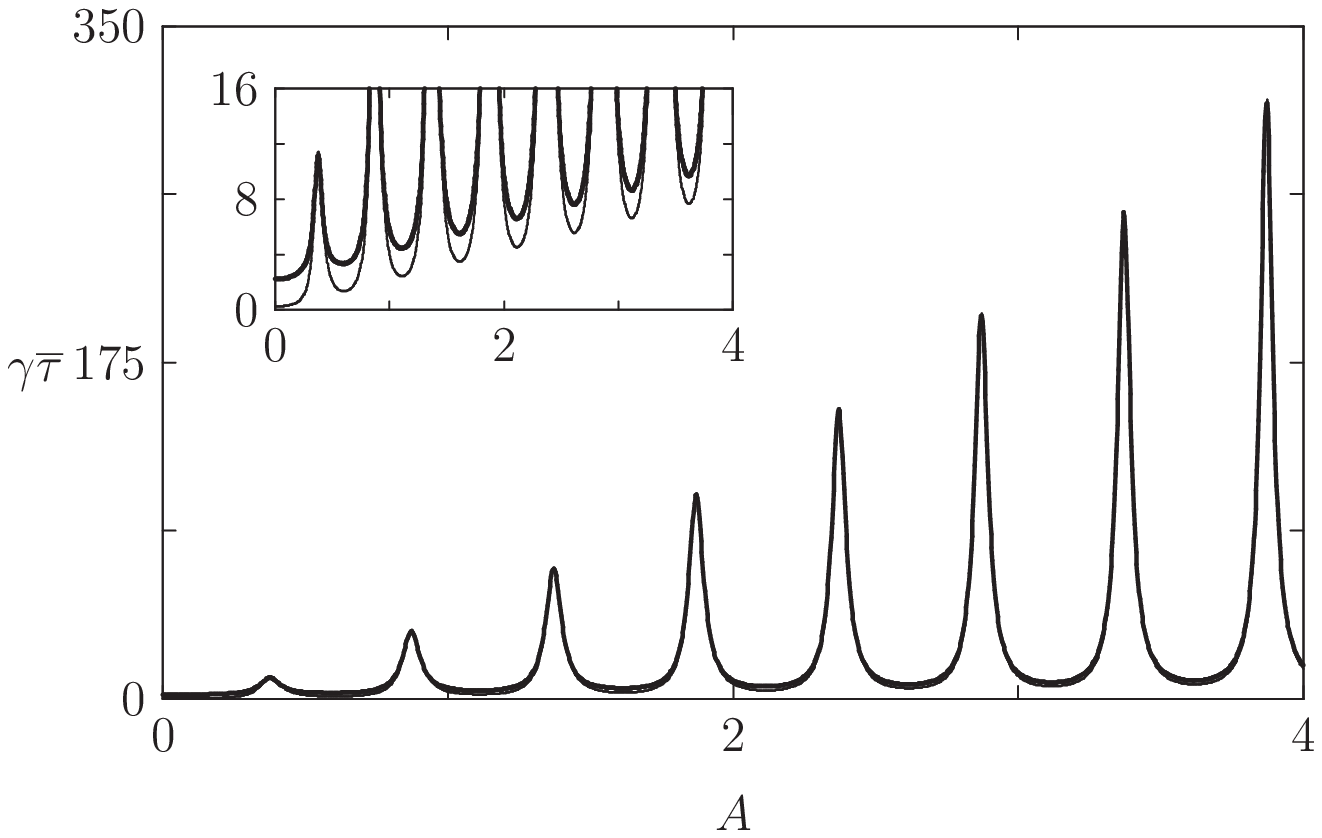,height=5cm}}

\vskip6pt

\hsize=5in
\parindent=1.5in
\baselineskip=6pt
\hang
{\eight Fig.~4. Exact (thick curve) and approximate [Eq.~(34)] (thin curve) semi-quantum mean waiting time
plotted as a function of the amplitude of oscillation in the trap.}}

\parindent=0.2in
\hsize=6.5in

\vskip6pt

{
\raggedright

\par
\vskip6pt
\noindent
\noindent{\it 4.2 The semi-quantum diffusion limit}:
\par
\vskip6pt
\noindent
Point (i) in Sec.~2 is concerned with the approach of the heating curves for small $\eta$ to a diffusion limit. It was
noted that the $\eta$-dependence in this limit amounts to a scaling of the time by $\eta^4$. We are now in a\break
position to elaborate upon these observations. We treat the diffusion limit for semi-quantum trajectories first, then
compare the description for quantum trajectories in the following section.

It is convenient to use a set of coordinates rotated to remove the phase of the complex amplitude $\tilde\alpha$.
In this frame of reference, the momentum ``kick'' defined by Eq.~(11b) is written as [$\zeta$ defined by Eq.~(23)]
$$
\mkern100mu\Delta(\theta,\phi,\zeta)=\Delta_\parallel(\theta,\phi,\zeta)+i\Delta_\perp(\theta,\phi,\zeta)=i\eta\sin\theta\cos\phi
e^{i\zeta},\eqno{(35)\hbox{\hskip2cm}}
$$
where $\Delta_\parallel(\theta,\phi,\zeta)$ is a ``kick'' in the amplitude of $\tilde\alpha$ and $\Delta_\perp(\theta,\phi,\zeta)$
is a ``kick'' in its phase. Successive ``kicks'' are distributed randomly, though not necessarily uniformly, over
the angles $\theta$, $\phi$, and $\zeta$. Taking an average over these distributions yields a covariance matrix
that depends only on the amplitude of\break oscillation in the trap $A$:
$$
\mkern100mu{\bf C}(A)={\bf B}(A){\bf B}^T\mkern-2mu(A)=\left(\matrix{\overline{\Delta_\parallel^2(\theta,\phi,\zeta)}&
\overline{\Delta_\parallel(\theta,\phi,\zeta)\Delta_\perp(\theta,\phi,\zeta)}\cr
\noalign{\vskip4pt}
\overline{\Delta_\parallel(\theta,\phi,\zeta)\Delta_\perp(\theta,\phi,\zeta)}&\overline{\Delta_\perp^2(\theta,\phi,\zeta)}}
\mkern3mu\right).\eqno{(36)\hbox{\hskip2cm}}
$$
The covariance matrix defines the r.m.s. size of the phase-space ``kicks''. It is plotted from a simulated trajectory
average in Fig.~5. To this information we add their average rate---i.e., the inverse of the mean waiting time plotted
in Fig.~4, also a function only of $A$. Assuming, then, that the phase-space ``kicks'', of order $\eta$ [Eq.~(35)],
are small, so that significant changes in $\tilde\alpha$ accumulate over very many photon scattering events, we may put
these pieces of information together to arrive at the stochastic differential equation with amplitude-dependent diffusion
$$
\mkern100mu{\bf R}[-\arg(\tilde\alpha)]\mkern-3mu\left(\matrix{d\tilde\alpha_\parallel\cr
\noalign{\vskip2pt}
d\tilde\alpha_\perp}\right)=\frac1{\sqrt{\bar\tau(A)}}\mkern3mu{\bf B}(A)\mkern-2mu\left(\matrix{dW_\parallel\cr
\noalign{\vskip2pt}
dW_\perp}\right),\eqno{(37)\hbox{\hskip2cm}}
$$
where the rotation matrix ${\bf R}[-\arg(\tilde\alpha)]$ brings the phase-space increments back to the original
coordinate system; $dW_\parallel$ and $dW_\perp$ are independent Wiener increments, with covariances
$$
\mkern100mu\overline{dW_\parallel dW_\parallel}=\overline{dW_\perp dW_\perp}=dt,\qquad\overline{dW_\parallel dW_\perp}=0.\eqno{(38)\hbox{\hskip2cm}}
$$

}

\vskip6pt

\vbox{\vskip0.5cm
\centerline{\psfig{figure=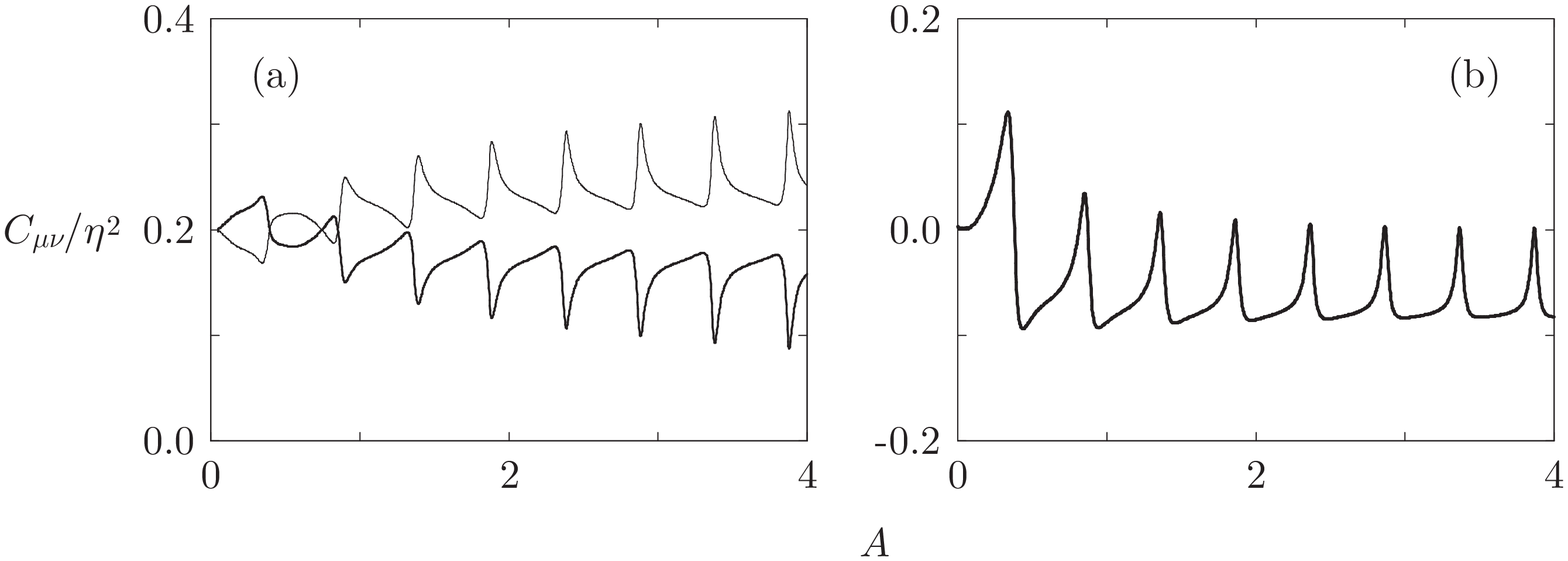,height=4.5cm}}

\vskip6pt

\hsize=5.75in
\parindent=0.75in
\baselineskip=6pt
\hang
{\eight Fig.~5. Semi-quantum covariance matrix elements plotted as a function of the amplitude of oscillation in the trap:
(a) amplitude (thick curve) and phase (thin curve) quadrature variances; (b) amplitude-phase cross-correlation.}
}

\parindent=0.2in
\hsize=6.5in

\vskip12pt

{
\raggedright

The noted $\eta^4$-scaling of the time axis in the diffusion limit [point (i) of Sec.~2] follows directly from Eq.~(35)--(37).
Note first that ${\bf B}(A)\sim\eta$, while, in addition, the oscillation amplitude in wavelengths is related to the
phase-space amplitude by $A=\eta|\tilde\alpha|/\pi$ [Eq.~(20)]; thus, after multiplying Eq.~(37) by $\eta/\pi$, the right-hand
side is of order $\eta^2$ and the fluctuation variance, at fixed $A$, grows as $\eta^4t$.

We find that the diffusion model agrees with semi-quantum jump simulations for Lamb-Dicke\break parameters on the order
of $\eta=0.2$ or less; in fact, there is very little difference for $\eta=0.4$ [results for\break $\eta=0.2$ and $0.4$ are
indistinguishable in Fig.~1(a)]. Further increase, however, brings noticeably faster semi-quantum heating, as shown
by the curves for $\eta=0.8$, $1.4$, $2.2$, and $2.6$ in Fig.~1(a) [point (iii) of Sec.~2]. In this regime the scale
of the individual quantum events (momentum ``kicks'') approaches that\break of the underlying phase-space structure represented
by the peaks in Fig.~4. A handful of jumps can\break destabilize the metastable amplitudes much faster than the diffusion
process can. 
\par
\vskip6pt
\noindent
\noindent{\it 4.3 Quantum trajectories compared}:
\par
\vskip6pt
\noindent
Semi-quantum trajectories are conceptually simple and provide a convenient starting point for our\break understanding. They
are nevertheless a rather gross approximation. Points (ii) and (iv) of Sec.~2 record two instances where the approximation
has a significant effect upon the heating rate. Considering first the quantitative difference in the diffusion limit
[point (ii)], we might ask how the mean waiting time and\break covariance matrix change as functions of the
amplitude of oscillation $A$. The mean waiting time computed from full quantum trajectories is plotted in Fig.~6. For
a Lamb-Dicke parameter of $\eta=0.1$ it is almost indistinguishable from the semi-quantum result. At larger values of
$\eta$, however, differences set in: there is a gradual smoothing out of the peaks, which is noticeable for $\eta=0.2$ and
virtually complete for $\eta=1.4$. This development is explained in Sec.~5.

Changes to the covariance matrix are more relevant for the diffusion (small $\eta$) limit. Here the most important point to
note is the difference between Eqs.~(11a) and (19a). The former specifies the semi-quantum phase-space ``kick'', while
the latter adds to this the local-frame displacement of the center-of-mass wavepacket, $\langle\hat a\rangle_{\rm local}$,
which occurs during the elapsed waiting time following the last quantum jump. We might refer to this as the
{\it stochastic dipole-force displacement\/}. It arises from the direct action of the standing-wave light field on the
center-of-mass of the ion. Since resonant excitation constitutes a\break nonperturbative interaction, it is not possible to
extract an optical potential to account for this.\break Nevertheless, to add to the momentum ``kicks'' from fluorescence,
there is also momentum exchanged\break between the standing-wave laser field and the ion; this the semi-quantum model omits,
while quantum\break trajectories include it as a time evolution of the center-of-mass kets---$|\bar\psi_{\rm REC}^{(-)}(t)\rangle$
and $|\bar\psi_{\rm REC}^{(+)}(t)\rangle$---taking place in between the quantum jumps [Eqs.~(17)].

}

\vskip6pt

\vbox{\vskip0.5cm
\centerline{\hskip-0.5cm\psfig{figure=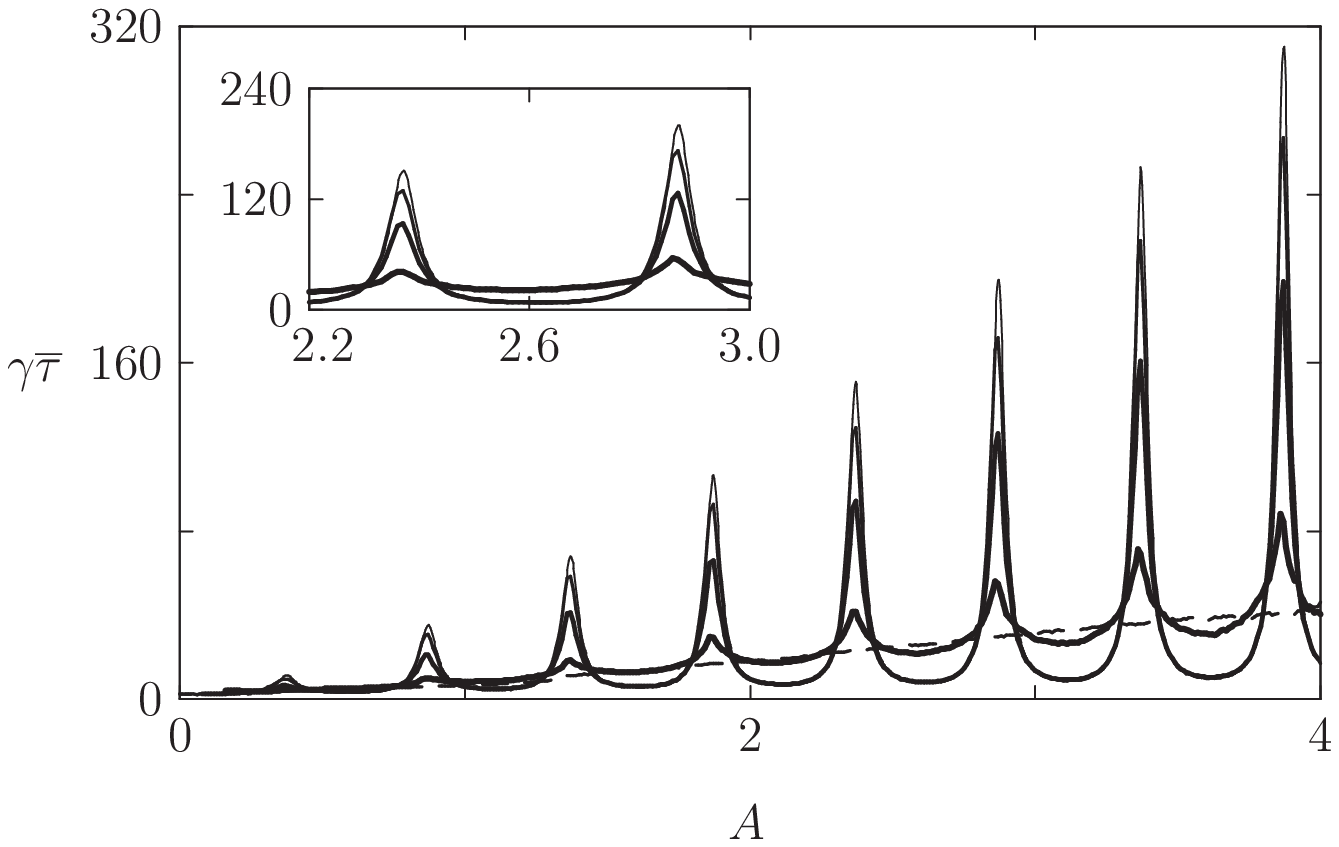,height=5cm}}

\vskip6pt

\hsize=5in
\parindent=1.5in
\baselineskip=6pt
\hang
{\eight Fig.~6. Mean waiting time from quantum trajectories plotted as a function of the amplitude of oscillation
in the trap; for Lamb-Dicke parameters \hbox{\eightmi\char17}$\mkern3mu$=$\mkern3mu$0.2, 0.4, 0.8 (thinner line to
thicker line), and 1.4 (dashed line). The tallest curve (thinest line) is the exact semi-quantum result plotted in Fig.~4.}}

\parindent=0.2in
\hsize=6.5in

\vskip6pt

{
\raggedright

From Eq.~(19b), when characterizing the individual quantum events for quantum trajectories, we\break replace Eq.~(35) by
$$
\mkern100mu\Delta(\theta,\phi,\zeta,\zeta^\prime)=|\langle\hat a\rangle_{\rm local}|e^{i\zeta^\prime}+i\eta\sin\theta\cos\phi
e^{i\zeta},\eqno{(39)\hbox{\hskip2cm}}
$$
where a dependence enters upon the additional relative phase
$$
\zeta^\prime=\arg(\langle\hat a\rangle_{\rm local})-\arg(\tilde\alpha).\eqno{(40)\hbox{\hskip2cm}}
$$
The covariance matrix is defined as before, with real and imaginary components of $\Delta(\theta,\phi,\zeta,\zeta^\prime)$
resolved\break in amplitude and phase directions, but now an average over the four phases $\theta$, $\phi$, $\eta$, $\eta^\prime$
is taken. The\break numerically computed result is displayed in Fig.~7. Its principal difference when compared to Fig.~5 is that
the phase variance peaks on the metastable amplitudes to reach a value an order of magnitude larger than in the semi-quantum case.
Other less obvious changes are observed; for example, the amplitude variance\break also increases to produce a broad background
near each metastable amplitude---the likely cause of the noted larger heating rate [point (ii) of Sec.~2].

The peaking of the phase variance on the metastable amplitudes appears most dramatically when\break phase-space plots are made.
Sample plots are shown in Fig.~8, where we plot the complex amplitude $\tilde\alpha$\break  from semi-quantum simulations (second
column in the figure) and $\langle\hat a\rangle_{\rm REC}$ from quantum simulations\break (third column). In the diffusion limit
($\eta=0.2$), the phase diffusion on the metastable amplitudes\break produces a dramatic ``bulls-eye'' pattern for quantum
trajectories (top right frame), in sharp contrast

}

\vskip6pt

\vbox{\vskip0.5cm
\centerline{\hskip-1cm\psfig{figure=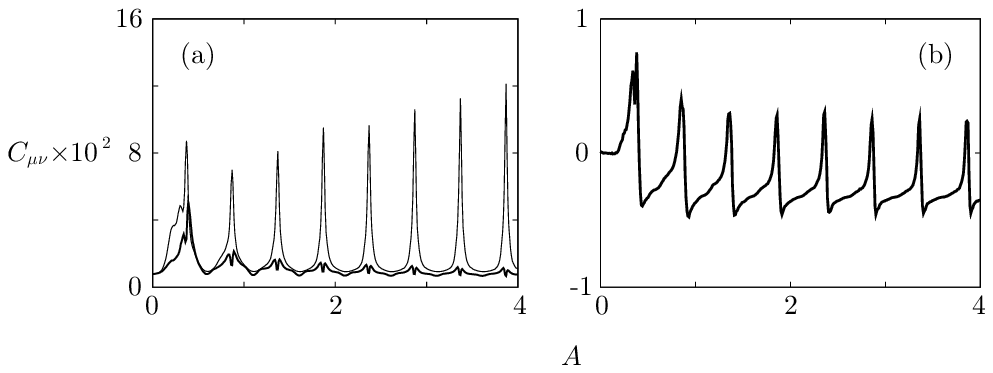,height=5.5cm}}

\vskip6pt

\hsize=5.75in
\parindent=0.7in
\baselineskip=6pt
\hang
{\eight Fig.~7. Full quantum covariance matrix elements plotted as a function of the amplitude of oscillation in the trap:
(a) amplitude (thick curve) and phase (thin curve) quadrature variances; (b) amplitude-phase cross-correlation. Results
computed as a Monte-Carlo average for \hbox{\eightmi\char17}$\mkern3mu$=$\mkern3mu$0.2; note that a small sampling error
remains.}
}

\parindent=0.2in
\hsize=6.5in

\vskip6pt

{
\raggedright

}

\vbox{\vskip0.5cm
\centerline{\psfig{figure=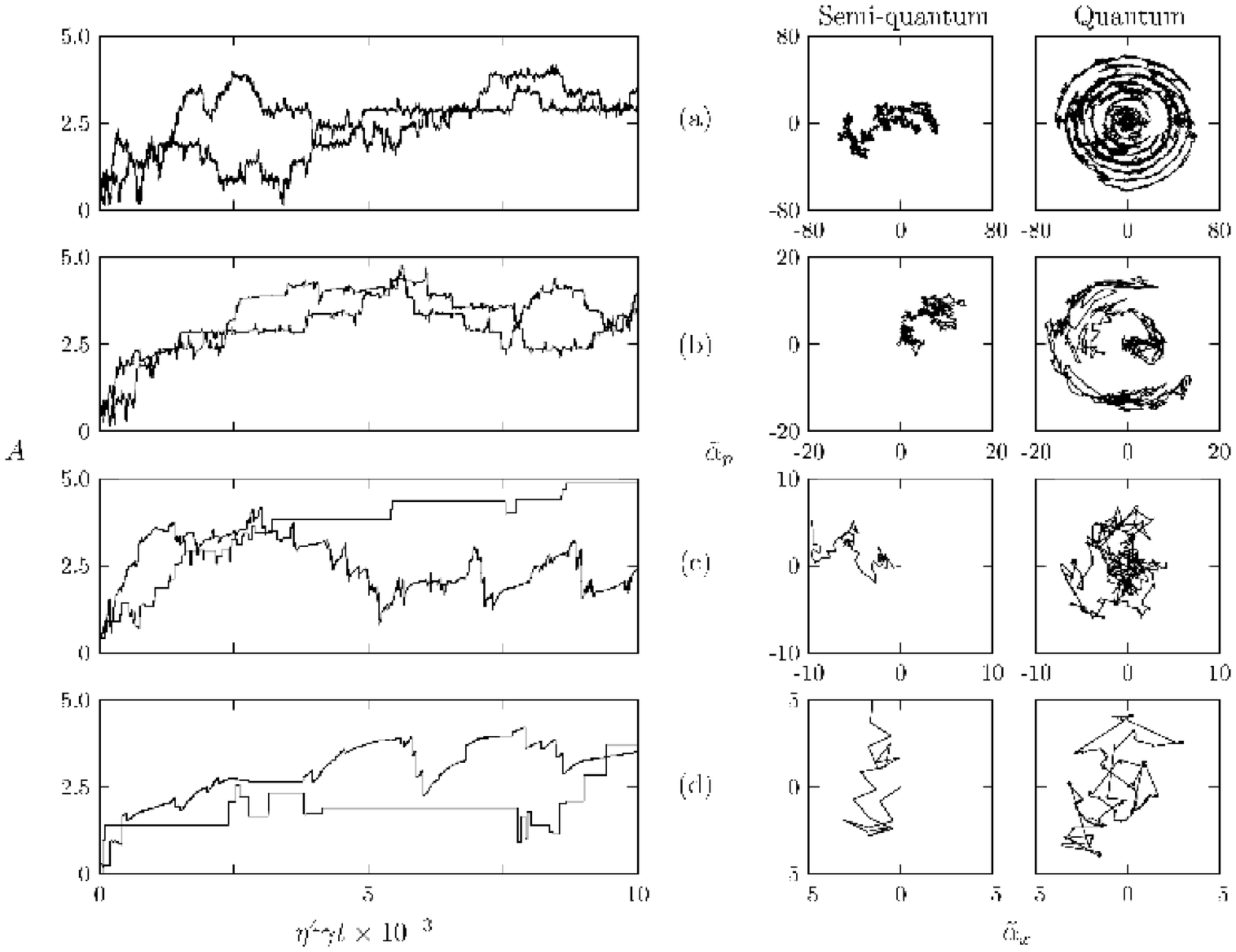,height=11cm}}

\vskip6pt

\hsize=6in
\parindent=0.5in
\baselineskip=6pt
\hang
{\eight Fig.~8. Sample semi-quantum (second column) and quantum (third column) trajectory simulations compared through plots
of the complex oscillator amplitude in phase space; for Lamb-Dicke parameters \hbox{\eightmi\char17}$\mkern3mu$=$\mkern3mu$0.2
(a), 0.8 (b), 1.4 (c), and 2.2 (d). The first column reproduces the plots from Fig.~2.}
}

\parindent=0.2in
\hsize=6.5in

\vskip12pt

{
\raggedright

\noindent
to the semi-quantum result. The phase diffusion is still in evidence for $\eta=0.8$ (second row), but gradually ceases to be
a feature as $\eta$ is further increased and the diffusion picture breaks down.

Our final comment on the comparison between semi-quantum and quantum trajectories for small\break Lamb-Dicke parameters is
illustrated by Fig.~9. Quantum trajectories provide a quantum description of the center-of-mass state---they evolve a
center-of-mass wavepacket. What, then, is the form of this state? Perhaps we can expect a good approximation to a minimum
uncertainty state, which is the closest\break quantum mechanical representation of the phase-space point of Eq.~(7).
As it turns out, this is not the case, even for the smallest Lamb-Dicke parameters considered (though for sufficiently
small $\eta$ we might still expect that result). The states shown as $Q$ functions in Fig.~9 were recovered for a Lamb-Dicke\break
parameter $\eta=0.2$. They illustrate a generally observed amplitude squeezing, with large squeezing in the vicinity of
the metastable amplitudes and significantly less squeezing at amplitudes in between. The\break degree of squeezing---of the
conditional (instantaneously sampled) wavepacket---is plotted as a function of the amplitude of oscillation in the trap
in the frame to the right in the figure.

We propose the following quantum measurement interpretation of these results. The ion fluorescence may be viewed as the record
of an imperfect measurement, one revealing information about the amplitude\break of oscillation in the trap. Most simply, if the
fluorescence rate is low, the ion likely oscillates with one of the metastable amplitudes; if high, with some amplitude midway
in between. Beyond this simplest\break statement, a more quantitative deduction about the actual amplitude of oscillation may
be made on the basis of the correlation between fluorescence rate and oscillation amplitude given in Figs.~4 and 6. The\break
important point to note is that the amplitude discrimination achieved is especially high in the vicinity of a metastable
amplitude and significantly poorer in between. Where the discrimination is high, the center-of-mass wavepacket becomes
squeezed so that its predicted uncertainty in the amplitude of oscillation is consistent with the data made available
though the measurement record (fluorescence)--i.e. the squeezing\break

}

\vbox{\vskip0.5cm
\centerline{\psfig{figure=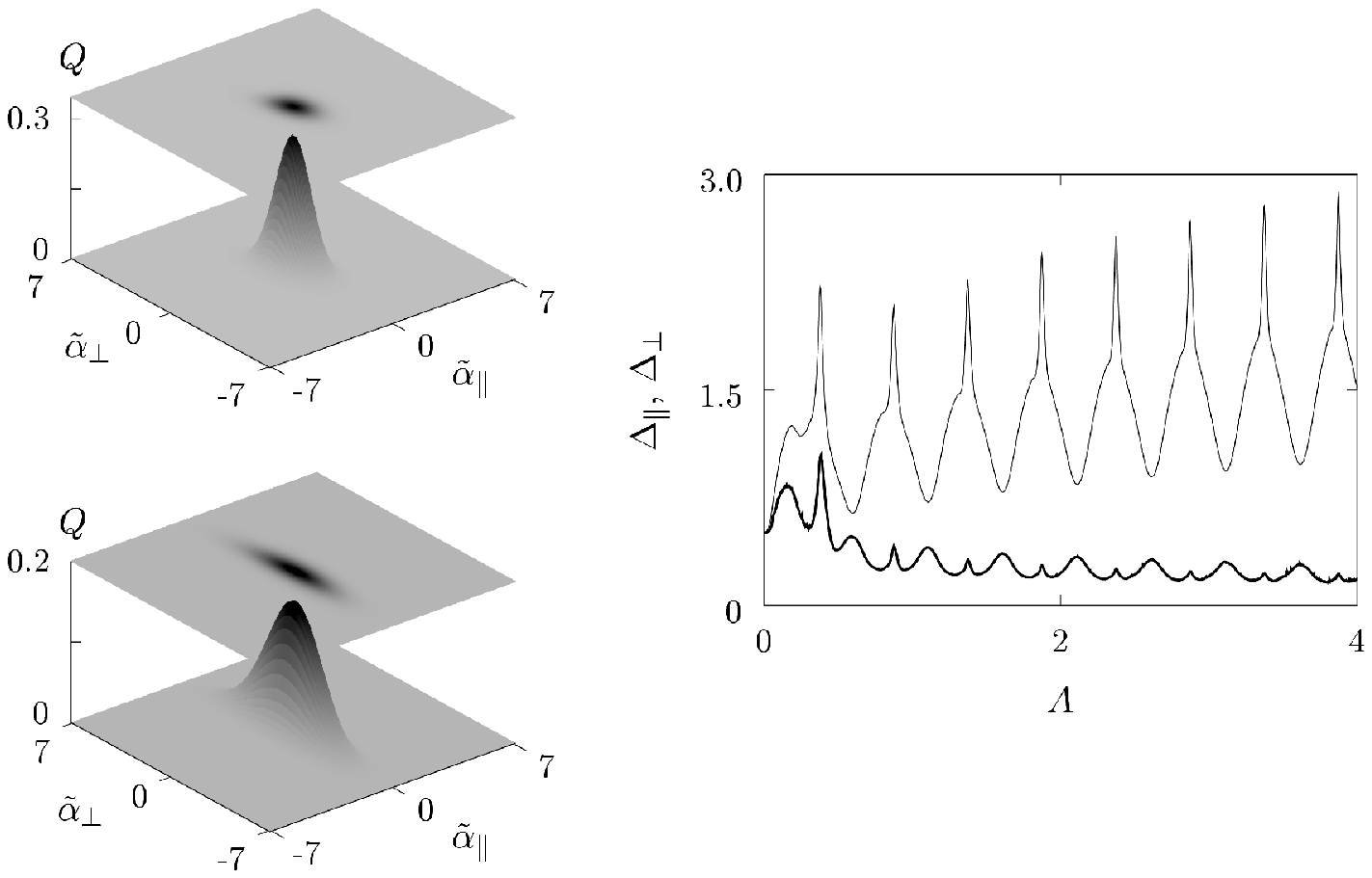,height=8cm}}

\vskip6pt

\hsize=6in
\parindent=0.5in
\baselineskip=6pt
\hang
{\eight Fig.~9. Squeezing of the center-of-mass wavepacket as a function of the amplitude of oscillation in the
trap. The scattered photons may be detected and read as the signal of an imperfect measurement of the amplitude
\hbox{\eightmi A}. Near a metastable amplitude, the discrimination of the measurement is very good and the
center-of-mass wavepacket becomes squeezed so as to be consistent with the record read (\hbox{\eightmi Q} function
on the lower left); in between the metastable amplitudes the discrimination, consequently the squeezing,
is not so good (\hbox{\eightmi Q} function on the upper left). To the right, amplitude (thick curve) and phase
(thin curve) quadrature variances are plotted as a function of the amplitude of oscillation in the trap. All results
are for a Lamb-Dicke parameter \hbox{\eightmi\char17}$\mkern3mu$=$\mkern3mu$0.2.}
}

\parindent=0.2in
\hsize=6.5in

\vskip12pt

{
\raggedright

\noindent
may be view as the product of a dynamical ``collapse of the wavepacket'', which in the case of a near-ideal
measurement would produce an amplitude eigenstate.

\vskip6pt
\noindent
{\tenb 5. Quantum inhibition of fluorescence and the limit of large quantum jumps}
\vskip6pt
\noindent
Points (iii) and (iv) of Sec.~2 note that large Lamb-Dicke parameters yield an $\eta$-dependence in the heating\break
curves of Fig.~1 over and above the $\eta^4$-scaling of the time axis---the dependence arising from the diffusion\break
limit. Point (iv) notes, in particular, that the quantum heating rate {\it decreases\/} for large $\eta$, while the\break
increase seen for semi-quantum trajectories seems more reasonable, when, due to their increased size, just a few
momentum ``kicks'' are sufficient to ``skip over'' (destabilize) the metastable amplitudes. In fact, the\break
lowered heating rate is caused by a suppression of fluorescence of an entirely different (to the mechanism of Sec.~4)
quantum mechanical origin, as we now explain.

Consider, for example, the matrix elements for excitation of the ion out of the ground electronic state and the
ground state of the trap while absorbing one photon and an even number of ``phonons''; the final state is
the excited electronic state and energy eigenstate state of the trap $|2n\rangle$, $n=0,1,2,\ldots$. The\break matrix
elements, as a function of $\eta$, are 
$$
\eqalignno{
\langle 2n|\cos[k\hat x(t)]|0\rangle&=\langle 2n|\cos\left[\eta\left(\hat ae^{-i\omega_Tt}+\hat a^\dagger
e^{i\omega_Tt}\right)\right]|0\rangle\cr
\noalign{\vskip2pt}
&=(-1)^n\frac{\eta^{2n}}{\sqrt{2n!}}\mkern2mu e^{-\eta^2/2}e^{i2n\omega_Tt}.&{(41)\hbox{\hskip2cm}\mkern-8mu}
}
$$
Note that matrix elements for the absorption of an odd number of ``phonons'' vanish due to the even\break parity of
$\cos(k\hat x)$. The squares of matrix elements (41) are plotted for the first few $n$ in Fig.~10(a). From the figure
we see that transitions $|0\rangle\to|0\rangle$ dominate in the diffusion limit. On the other hand, for $\eta\sim1$,
the ground-state wavepacket extends beyond $\pm\lambda/4$ where the cosine function changes sign. With the sign change
the value of the matrix element is reduced. Thus, at $\eta=1.5$, the $|0\rangle\to|2\rangle$ transition is strongest,
while for $\eta\sim2$, $|0\rangle\to|4\rangle$ and $|0\rangle\to|6\rangle$ are the strongest transitions. We note then
that these strongest, 

}

\vskip6pt

\vbox{\vskip0.5cm

\centerline{\hskip-0.5cm\psfig{figure=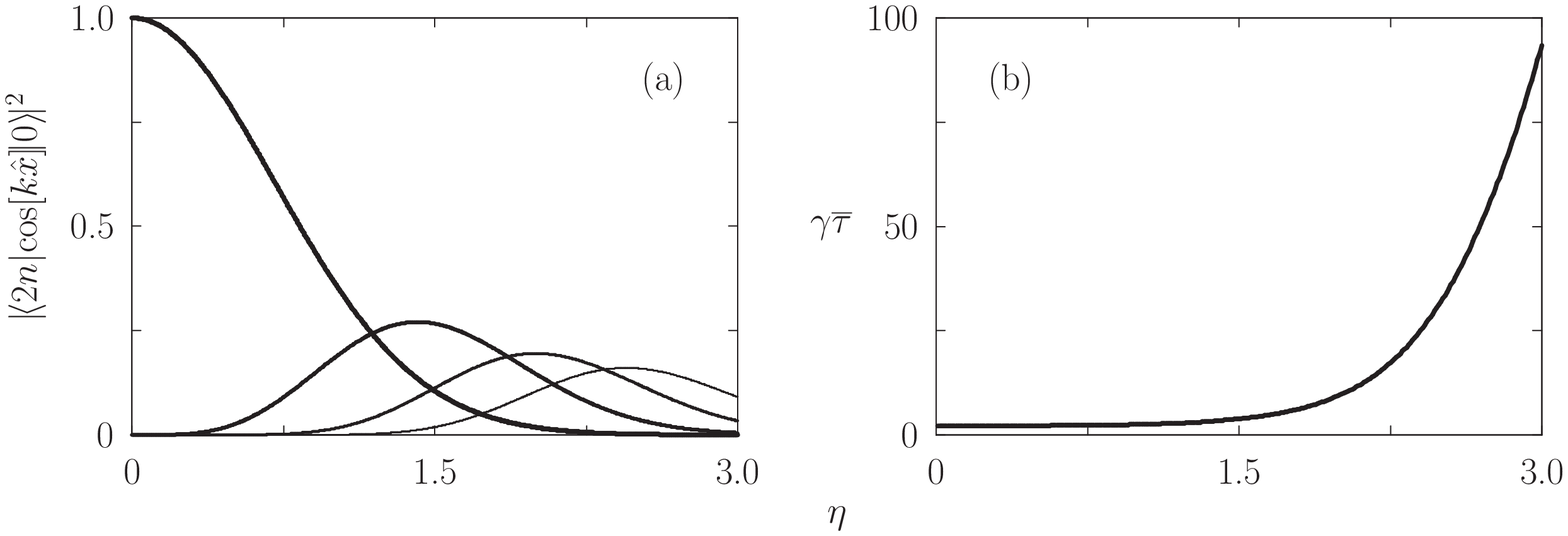,height=4.5cm}} 

\vskip6pt

\hsize=5.75in
\parindent=0.75in
\baselineskip=6pt
\hang
{\eight Fig.~10. (a) Matrix elements for transitions out of the ground state of the trap,
\hbox{\eightsy\char106}0\hbox{\eightsy\char105}, to the excited state
\hbox{\eightsy\char106}2\hbox{\eightmi n}\hbox{\eightsy\char105}, plotted as a function of Lamb-Dicke parameter
\hbox{\eightmi\char17} for \hbox{\eightmi n}$\mkern3mu$=$\mkern3mu$0, 1, 2, and 3
(thicker curve to thinner curve); (b) Mean waiting time until the first quantum jump as a function
\hbox{\eightmi\char17}.}}

\parindent=0.2in
\hsize=6.5in

\vskip6pt

{
\raggedright

\noindent
multi-phonon transitions are not excited on resonance; the $|0\rangle\to|2n\rangle$ transition is detuned by an\break amount
$2n\omega_T=2n\gamma$ [Eq.~(5)]. The combined effect of these two observations leads to a strong quantum\break suppression
of the fluorescence; hence the reduced heating rate of Fig.~1(b). As an illustration of the\break suppression, in Fig.~10(b)
we plot, as a function of $\eta$, the mean waiting time---beginning in the ground state of the trap---for the very first
photon to be scattered. The waiting time grows rapidly for Lamb-Dicke parameters larger than $\eta=1.5$.

There is a great deal more to be said about the quantum-stochastic motion of a trapped ion in the regime of large Lamb-Dicke
parameters. This regime is similar to that of optical frequency cavity QED, where the influence of single scattering events
is no longer small and the diffusion picture fails, or becomes forced at best [32]. As an indication of what is on offer,
we conclude with Fig.~11. Here we present\break 

}

\vbox{\vskip0.5cm
\centerline{\psfig{figure=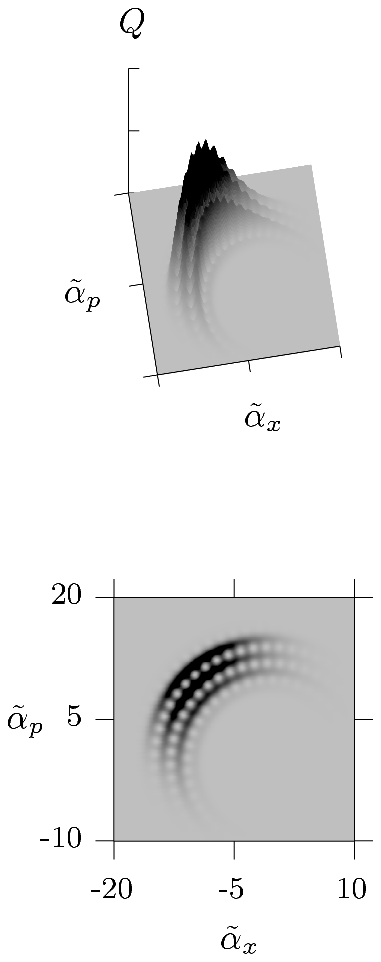,height=10cm}\raise0.3cm\hbox{\psfig{figure=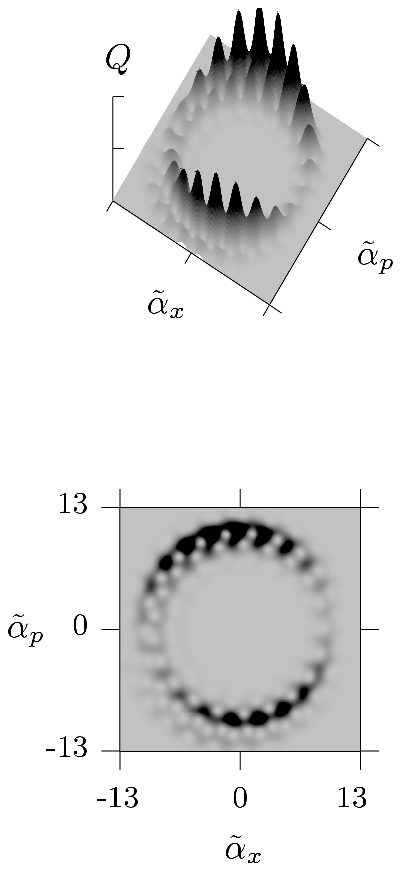,height=8.75cm}}\hskip0.8cm\psfig{figure=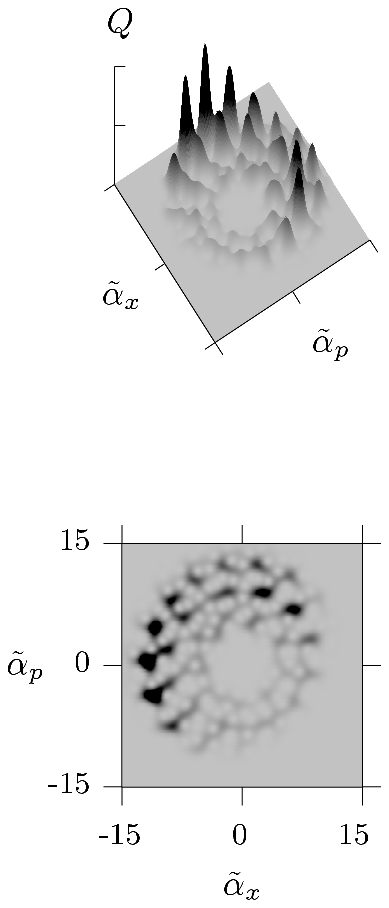,height=9.46cm}}

\vskip6pt

\hsize=6in
\parindent=0.5in
\baselineskip=6pt
\hang
{\eight Fig.~11. Sample states from quantum trajectories for Lamb-Dicke parameters \hbox{\eightmi\char17}=0.8 (left plots)
1.4 (center plots), and 2.2 (right plots). \hbox{\eightmi Q} functions for the conditional center-of-mass state are plotted.}
}

\parindent=0.2in
\hsize=6.5in

\vskip12pt

\raggedright

\noindent
$Q$ functions for sample states reached with Lamb-Dicke parameters $\eta=0.8$, $1.4$, and $2.2$. Of course, a huge variety
of such states are visited throughout the course of a quantum trajectory; the question of what\break interesting quantities
might be easily measured has yet to be explored. Note, however, the progression from the amplitude squeezed state of Fig.~9
to the states with increasing phase uncertainty of Fig.~11. Note also the interesting development of the amplitude uncertainty.
The state for $\eta=0.8$ (on the left in the figure), for example, resides on three metastable amplitudes at once, as a
superposition. Then there is the development from left to right---with increasing $\eta$---where, for $\eta=2.2$, a somewhat
disordered phase-space structure emerges from the quantum interference pattern. Clearly, there remains much interesting behavior
to be investigated in future work.

\vskip12pt
\noindent
Work supported by the Marsden fund of the RSNZ.

\vskip12pt
\noindent
{\tenb 6. References}

\vskip6pt
\noindent
{\eight [1] W. Nagourney, J. Sandberg, and H. Dehmelt, ``Shelved Optical Electron Amplifier: Observation of Quantum Jumps," Phys.\ Rev.\ Lett.}\ {\eightb 56}{\eight , 2797 (1986)}.
\par
\vskip4pt
\noindent
{\eight [2] Th. Sauter, W. Neuhauser, R. Blatt, and P.E. Toschek, ``Observation of Quantum Jumps,'' Phys.\ Rev.\ Lett.}\ {\eightb 57}{\eight , 1696 (1986)}.
\par
\vskip4pt
\noindent
{\eight [3] J.C. Berquist, R.O. Hulet, W.M. Itano, and D.J. Wineland, ``Observation of Quantum Jumps in a Single Atom,'' Phys.\ Rev.\ Lett.}\ {\eightb 57}{\eight , 1699 (1986)}.
\par
\vskip4pt
\noindent
{\eight [4] S. Stenholm, ``The semiclassical theory of laser cooling,'' Rev.\ Mod.\ Phys.}\ {\eightb 58}{\eight , 699 (1986)}.
\par
\vskip4pt
\noindent
{\eight [5] D.J. Wineland and W.M. Itano, ``Laser cooling,'' Physics Today} {\eightb 40}{\eight(6), 34 (1987)}.
\par
\vskip4pt
\noindent
{\eight [6] C. Cohen-Tannoudji, ``Laser cooling and trapping of neutral atoms: theory,'' Phys.\ Rep.}\ {\eightb 219}{\eight , 153 (1992)}.
\par
\vskip4pt
\noindent
{\eight [7] F. Diedrich, J.C. Bergquist, W.M. Itano, and D.J. Wineland, ``Laser Cooling to the Zero-Point Energy of Motion,'' Phys.\ Rev.\ Lett.}\ {\eightb 62}{\eight , 403 (1989)}.
\par
\vskip4pt
\noindent
{\eight [8] C. Monroe, D.M. Meekof, B.E. King, S.R. Jefferts, W.M. Itano, and D.J. Wineland, ``Resolved-Sideband Raman Cooling of a Bound Atom to the 3D Zero-Point Energy,'' Phys.\ Rev.\ Lett.}\ {\eightb 75}{\eight , 4011 (1995)}.
\par
\vskip4pt
\noindent
{\eight [9] D. Leibfried, R. Blatt, C. Monroe, and D.J. Wineland, ``Quantum dynamics of single trapped ions,'' Rev.\ Mod.\ Phys.}\ {\eightb 75}{\eight , 281 (2003)}.
\par
\vskip4pt
\noindent
{\eight [10] J.T. H\"offges, H.W. Baldauf, T. Eichler, S.R. Helmfrid, and H. Walther, ``Heterodyne measurement of the fluorescent radiation of a single trapped ion,'' Optics Commun.}\ {\eightb 133}{\eight , 1701 (1997)}.
\par
\vskip4pt
\noindent
{\eight [11] J.T. H\"offges, H.W. Baldauf, W. Lange, and H. Walther, ``Heterodyne measurement of the resonance fluorescence of a single ion,'' J. Mod.\ Opt.}\ {\eightb 44}{\eight , 1999 (1997)}.
\par
\vskip4pt
\noindent
{\eight [12] Ch.\ Raab, J. Eschner, J. Bolle, H. Oberst, F. Schmidt-Kaler, and R. Blatt., ``Motional Sidebands and Direct Measurement of the Cooling Rate in the Resonance Fluorescence of a Single Trapped Ion,'' Phys.\ Rev.\ Lett.}\ {\eightb 85}{\eight , 538 (2000)}.
\par
\vskip4pt
\noindent
{\eight [13] F. Diedrich and H. Walther, ``Nonclassical Radiation of a Single Stored Ion,'' Phys.\ Rev.\ Lett.}\ {\eightb 58}{\eight , 203 (1987)}.
\par
\vskip4pt
\noindent
{\eight [14] W.M. Itano, J.C. Bergquist, and D.J. Wineland, Photon antibunching and sub-Poissonian statistics from quantum jumps in one and two atoms,'' Phys.\ Rev.\ A} {\eightb 38}{\eight , 559(R) (1988)}.
\par
\vskip4pt
\noindent
{\eight [15] M. Schubert, I. Siemers, R. Blatt, W. Neuhauser, and P.E. Toschek, ``Photon Antibunching and Non-Poissonian Fluorescence of a Single Three-Level Ion,'' Phys.\ Rev.\ Lett.}\ {\eightb 68}{\eight , 3016 (1992)}.
\par
\vskip4pt
\noindent
{\eight [16] D.M. Meekhof, C. Monroe, B.E. King, W.M. Itano, and D.J. Wineland, ``Generation of Nonclassical Motional States of a Trapped Atom,'' Phys.\ Rev.\ Lett.}\ {\eightb 76}{\eight , 1796 (1996)}.
\par
\vskip4pt
\noindent
{\eight [17] C. Monroe, D.M. Meekhof, B.E. King, and D.J. Wineland, ``A `Schr\"odinger Cat' Superposition State of an Atom,'' Science} {\eightb 272}{\eight , 1131 (1996)}.
\par
\vskip4pt
\noindent
{\eight [18] Ch.\ Roos, Th.\ Zeiger, H. Rohde, H.C. N\"agerl, J. Eschner, D. Leibfried, F. Schmidt-Kaler, and R. Blatt, Quantum State\break Engineering on an Optical Transition and Decoherence in a Paul Trap,'' Phys.\ Rev.\ Lett.}\ {\eightb 23}{\eight , 4713 (1999)}.
\par
\vskip4pt
\noindent
{\eight [19] C.J. Myatt, B.E. King, Q.A. Turchette, C.A. Sackett, D. Kielpinski, W.M. Itano, and D.J. Wineland, ``Decoherence of\break quantum superpositions through coupling to engineered reservoirs,'' Nature (London)} {\eightb 403}{\eight , 269 (2000)}.
\par
\vskip4pt
\noindent
{\eight [20] Q.A. Turchette, C.J. Myatt, B.E. King, C.A. Sackett, D. Kielpinski, W.M. Itano, C. Mmonroe, and D.J. Wineland,\break ``Decoherence and decay of motional quantum states of a trapped atom coupled to engineered reservoirs,'' Phys.\ Rev.\ A} {\eightb 62}{\eight , 053807 (2000)}.
\par
\vskip4pt
\noindent
{\eight [21] D. Leibfried, E. Knill, S. Seidelin, J. Britton, R.B. Blakestad, J. Chiaverini, D.B. Hume,W.M. Itano, J.D. Jost, C. Langer,\break R. Ozeri, R. Reichle, and D. J. Wineland, Creation of a six-atom `Schr\"odinger cat' state,'' Nature} {\eightb 438}{\eight , 639 (2005)}.
\par
\vskip4pt
\noindent
{\eight [22] H. H\"affner, W. H\"ansel, C.F. Roos, J. Benhelm, D. Chek-al-kar, M. Chwalla, T. K\"orber, U.D. Rapol, M. Riebe, P.O. Schmidt, C. Becher, O. G\"uhne, W. D\" ur, and R. Blatt, ``Scalable multiparticle entanglement of trapped ions,''  Nature} {\eightb 438}{\eight , 643 (2005)}.
\par
\vskip4pt
\noindent
{\eight [23] S. Seidelin, J. Chiaverini, R. Reichle, J.J. Bollinger, D. Leibfried, J. Britton, J.H. Wesenberg, R.B. Blakestad, R.J. Epstein, D.B. Hume, W.M. Itano, J.D. Jost, C. Langer, R. Ozeri, N. Shiga, and D.J. Wineland, ``Microfabrication Surface-Electrode Ion Trap for Scalable Quantum Information Processing,'' Phys.\ Rev.\ Lett.}\ {\eightb 96}{\eight , 253003 (2006)}.
\par
\vskip4pt
\noindent
{\eight [24] W. Neuhauser, M. Hohenstatt, P. Toschek, and H. Dehmelt, ``Optical-Sideband Cooling of Visible Atom Cloud Confined in a Parabolic Well,'' Phys.\ Rev.\ Lett.}\ {\eightb 41}{\eight , 233 (1978)}.
\par
\vskip4pt
\noindent
{\eight [25] D.J. Wineland, J. Dalibard, and C. Cohen-Tannoudji, ``Sisyphus cooling of a bound atom,'' J. Opt.\ Soc.\ Am.\ B} {\eightb 9}{\eight , 31 (1992)}.
\par
\vskip4pt
\noindent
{\eight [26] J.I. Cirac, R. Blatt, P. Zoller, and W.D. Phillips, ``Laser cooling of trapped ions in a standing wave,'' Phys.\ Rev.\ A} {\eightb 46}{\eight , 2668 (1992)}.
\par
\vskip4pt
\noindent
{\eight [27] J.I. Cirac and P. Zoller, ``Quantum Computations with Cold Trapped Ions,'' Phys. Rev. Lett.}\ {\eightb 74}{\eight , 4091 (1995)}.
\par
\vskip4pt
\noindent
{\eight [28] C. Monroe, D.M. Meekhof, B.E. King, W.M. Itano, and D.J. Wineland, ``Demonstration of a Fundamental Quantum Logic Gate,'' Phys. Rev. Lett.}\ {\eightb 75}{\eight , 4714 (1995)}.
\par
\vskip4pt
\noindent
{\eight [29] J. Javanainen, ``Light-induced motion of trapped ions I: low-intensity limit,'' J. Phys.\ B: Atom.\ Mol.\ Phys.}\ {\eightb 14}{\eight , 2519 (1981)}.
\par
\vskip4pt
\noindent
{\eight [30] J. Javanainen, ``Light-induced motion of trapped ions: III. Expansion around the recoilless solution,'' J. Phys.\ B: Atom.\ Mol.\ Phys.}\ {\eightb 18}{\eight , 1549 (1985)}.\par
\vskip4pt
\noindent
{\eight [31] J. Dalibard and C. Cohen-Tannoudji, ``Atomic motion in laser light: connection between semiclassical and quantum\break descriptions,'' J. Phys.\ B:Atom.\ Mol.\ Phys.}\ {\eightb 18}{\eight , 1661 (1985)}.
\vskip4pt
\noindent
{\eight [32] J. Dalibard and C. Cohen-Tannoudji, ``Dressed-atom approach to atomic motion in laser light: the dipole force revisited,''\break J. Opt.\ Soc.\ Am.\ B} {\eightb 2}{\eight , 1707 (1985)}.
\vskip4pt
\noindent
{\eight [33] H.J. Carmichael,} {\eighti Statistical Methods in Quantum Optics 2:~Non-classical Fields~}{\eight(Springer-Verlag, Berlin, 2007), Chaps.~12--16}.
\vskip4pt
\noindent
{\eight [34] H.J. Carmichael,} {\eighti An Open Systems Approach to Quantum Optics}{\eight, Lecture Notes in Physics, New Series m -- Monographs, Vol.~m18 (Springer-Verlag, Berlin, 1993), Chaps.~7--10}.
\vskip4pt
\noindent
{\eight [35] J. Dalibard, Y. Castin, and K. M\o lmer,``Wave-function Approach to Dissipative Processes in Quantum Optics,'' Phys.\ Rev.\ Lett.} {\eightb 68}{\eight , 580 (1992)}.
\vskip4pt
\noindent
{\eight [36] R. Dum, P. Zoller, and H. Ritsch,``Monte Carlo simulation of the master equations for spontaneous emission,'' Phys.\ Rev.\ A} {\eightb 45}{\eight , 4879 (1992)}.
\vskip4pt
\noindent
{\eight [37] K. M\o lmer, Y. Castin, and J. Dalibard,``Monte Carlo wave-function method in quantum optics,'' J. Opt.\ Soc.\ Am.\ B} {\eightb 10}{\eight , 524 (1993)}.
\vskip4pt
\noindent
{\eight [38] Y. Castin and K. M\o lmer,``Monte Carlo Wave-function Analysis of 3D Optical Molasses,'' Phys.\ Rev.\ Lett.}\ {\eightb 74}{\eight , 3772 (1995)}.
\vskip4pt
\noindent
{\eight [39] H.J. Carmichael, S. Singh, R. Vyas, and P.R. Rice,``Photoelectron waiting times and atomic state reduction in resonance fluorescence,'' Phys.\ Rev.\ A} {\eightb 39}{\eight , 1200 (1989)}.
\vskip4pt
\noindent
{\eight [40] M. Abramowitz and I.A. Stegun,} {\eighti Handbook of Mathematical Functions~}{\eight(Dover, New York, 1965), \S 9.1.44, p.361}.
\bye